\documentclass[aps,pre,amsmath,amssymb,floatfix,nofootinbib,showkeys, preprint]{revtex4-2}
\usepackage{graphicx} 
\usepackage{caption}
\usepackage{amsmath}
\usepackage{amsfonts}

\begin{document}
\title{Identifying Influential and Vulnerable Nodes in Interaction Networks through Estimation of Transfer Entropy Between Univariate and Multivariate Time Series}
\author{Julian Lee}
\email{jul@ssu.ac.kr}
\affiliation{Department of Bioinformatics and Life Science, Soongsil University, Seoul, Korea}
\date{\today}
\begin{abstract}
Transfer entropy (TE) is a powerful tool for measuring causal relationships within interaction networks. Traditionally, TE and its conditional variants are applied pairwise between dynamic variables to infer these causal relationships. However, identifying the most influential or vulnerable node in a system requires measuring the causal influence of each component on the entire system and vice versa. In this paper, I propose using outgoing and incoming transfer entropy---where outgoing TE quantifies the influence of a node on the rest of the system, and incoming TE measures the influence of the rest of the system on the node. The node with the highest outgoing TE is identified as the most influential, or ``hub", while the node with the highest incoming TE is the most vulnerable, or ``anti-hub". Since these measures involve transfer entropy between univariate and multivariate time series, naive estimation methods can result in significant errors, particularly when the number of variables is comparable to or exceeds the number of samples. To address this, I introduce a novel estimation scheme that computes outgoing and incoming TE only between significantly interacting partners. The feasibility of this approach is demonstrated using synthetic data and by applying it to real data of oral microbiota. The method successfully identifies the bacterial species known to be key players in the bacterial community, demonstrating the power of the new method.
\end{abstract}
\maketitle
\newpage
\section{Introduction}
Causal inference in interacting dynamic systems is a critical area of study that aims to understand the cause-and-effect relationships within complex systems~\cite{TE00, Pe00, Sp00, stan08, vejm08, ay08, vice11, wibr13, jae1, jae2, runge12a, runge12b,  sun14, sun15, runge15a, runge15b, runge18, runge19, lizi19,NL21,nove20}. These systems can be found in various fields, including neuroscience, economics, and ecology. The goal is to distinguish true causal interactions from mere correlations, which is essential for predicting system behavior and designing effective interventions.

 Information theory offers rigorous foundation for causal inference by quantifying the information shared between a pair of variables $X$ and $Y$ through mutual information $I(X,Y)$ defined as:
\begin{equation}
    I (X,Y) \equiv \left\langle \log_2 \left( \frac{ P(X,Y)}{P(X) P(Y)} \right)  \right\rangle
\end{equation}
where $P(X)$ and $P(Y)$ are the marginal probability distributions for the random variables $X$ and $Y$, $P(X,Y)$ are their joint probability distribution, and $\langle \rangle$ denote the expectation value.  In systems with more than two variables, one often seeks the direct correlation between $X$ and $Y$ after accounting for other variables. Denoting all variables other than $X$ and $Y$ by $Z$, the conditional mutual information $I(X,Y/Z)$ is  defined as:
\begin{equation}
    I (X,Y/Z) \equiv \left\langle \log_2 \left( \frac{P(X,Y/Z)}{P(X/Z) P(Y/Z)}\right) \right\rangle,
\end{equation}
which quantifies the direct correlation between $X$ and $Y$.

In dynamic systems, given two time series $X(t)$ and $Y(t)$, where the integer $t$ is an index for a discretized time, one might initially attempt to quantity the causal influence of $X$ on $Y$ using $I(X_-, Y(t))$, where $X_- \equiv (X(1), X(2), \cdots X(t-1))$ represents the past history of $X$. However, as previously mentioned, $I(X_-,Y(t))$  only quantifies the shared information between the history of $X$ and the present state of $Y$, not the causal influence of $X$ on $Y$. It is possible that $Y$ actually  causes $X$, which could still result in a non-zero value of $I(X_-, Y(t))$. Therefore, to quantify the true causal influence of $X$ on $Y$, we must correct for the effect due to the history of $Y$, using the measure:
\begin{equation}
    T_{X \to Y} \equiv {I}(X_-, Y(t)/ Y_-). \label{te1}
\end{equation}
This is known as transfer entropy (TE) from $X$ to $Y$~\cite{TE00}. Transfer entropy can also be expressed as:
\begin{equation}
        T_{X \to Y} =  H(Y(t) /  Y_- )   - H(Y(t) /  X_-,Y_- ) \label{te2}, 
\end{equation}
where
\begin{eqnarray}
         H (Y(t) /  Y_- ) &\equiv& \langle -\log_2 P(Y(t)/Y_-) \rangle  \nonumber\\
        H (Y(t) /  X_-,Y_- ) &\equiv&   \langle - \log_2 P(Y(t)/X_-, Y_-) \rangle \label{ce}. 
\end{eqnarray}
Here, the conditional entropies $H (Y(t) /  Y_- )$ and $ H (Y(t) /  X_-,Y_- )$ represent the uncertainties of $Y(t)$ after observing the history of $Y$ and after observing the histories of both $X$ and $Y$, respectively. Thus, $T_{X \to Y}$ can be interpreted as the reduction in the uncertainty of $Y(t)$ after observing the history of $X$, given that we already know the history of $Y$. It can be proved that $T_{X \to Y}$ as well as $H(Y(t) / Y_-)$ and $H(Y(t) / X_-,Y_-)$ are all non-negative~(see Appendex A). Specifically, this implies that if $Y(t)$ is completely determined by $Y_-$ such that $H(Y(t)/Y_-)=0$, then $T_{X \to Y}=0$. The intuitive interpretation is clear: If there is no uncertainty on $Y(t)$ after we observe $Y_-$, then obviously there is no additional uncertainty to be removed by knowing $X_-$.  

In systems with more than two time series, we are often interested in the direct causal influence of $X$ on $Y$ after accounting for indirect effects due to other variables. Denoting the multivariate time series of variables other than $X$ and $Y$ by $Z$, the direct causal influence of $X$ on $Y$ is quantified by the measure~\cite{runge12a, runge12b, sun14, sun15}
\begin{equation}
   T_{(X \to Y)/Z}  \equiv I(X_-, Y(t)/ Y_-, Z_-).
\end{equation}
This measure is called multivariate transfer entropy, conditional transfer entropy, causation entropy, or full conditional mutual information~\cite{runge12a, runge12b, sun14, sun15,  runge15a, runge15b, runge18, runge19, lizi19,NL21}. 

Although there are nuances in interpreting transfer entropy and its conditional variant as information flow or causal influence quantification~\cite{tecrit}, these measures have been widely used to uncover causal relationships in complex networks, including neural networks~\cite{neu1,neu2,neu3}, social networks~\cite{soc1}, and gene regulatory networks~\cite{tenet,scmte}.

However, in all these applications, the causal relationship between each pair of variables in the network was estimated. In this work, I focus on finding the most influential components in the network, the hubs, and most vulnerable components, which I will refer to as ``anti-hubs". Identifying hubs and anti-hubs is crucial for understanding, managing, and optimizing complex systems across various domains, from biology and ecology to engineering and social sciences. To find such influential or vulnerable components, we must quantify the causal influence of each node on the rest of the system and vice versa. This can be achieved by measuring the transfer entropy from the node of interest to the rest of the system,  referred to as  ``outgoing TE (OutTE)", and the transfer entropy  from the rest of the system to the node, referred to as ``incoming TE (InTE)", respectively. Here, we only have two variables in the system so that we can use Eq.(\ref{te1}), but with the source variable being univariate and the target variable being multivariate, or vice versa. As will be elaborated further, naive estimates of such quantities can lead to significant estimation errors, particularly when the number of variables is comparable to or larger than the length of the time series. In this work, I will introduce the novel estimation method for the OutTE and InTE, where estimation is performed only between significantly interacting partners.

The outline of the paper is as follows. In section II, I will illustrate the estimation problem of outTE and inTE using synthetic data from simple models, showing how  estimation error increases as the number of unrelated nodes increases. In section III, I will introduce the novel estimation method for the OutTE and InTE, which overcomes the estimation problem. In section IV, I will apply my method to microbiota data, showing that the new estimation method successfully identifies the bacterial species known to be key players in the bacterial community. The section V concludes the paper. 

\section{Estimation problem of Outgoing and Incoming Transfer entropy.}
The focus of this work is on identifying hub and anti-hub nodes in an interaction network: nodes that exert significant influence on the rest of the system, and those that are most influenced by it, respectively. To identify hub nodes, we need to compute OutTE for each node $X$, ${\rm OutTE}(X) \equiv T_{X \to {\rm rest}}$, where ``rest" denotes the collection of all 
 nodes except $X$. We then select the nodes with the largest values of ${\rm OutTE}$. Similarly, anti-hubs are identified by computing ${\rm InTE}(X) \equiv T_{{\rm rest} \to X}$ and selecting the nodes with the largest values of ${\rm InTE}$. However, in practice, the true values of these quantities are not available and must be {\it estimated} from data, which can introduce estimation errors.
 
 For example, consider microbiota data obtained from the saliva of a subject observed over 226 days~\cite{2series}, which will be discussed further in a later section. There are 879 nodes, where each node represents an operational taxonomic unit (OTU) in the microbiota. Transfer entropy from each of the node to the remaining 878 nodes can be estimated using a publicly available tool~\cite{jidt}, and the result of such an estimate is zero for all of the 879 nodes! This suggests that the estimate $\widehat {\rm OutTE}(X)$ is much smaller than the true value ${\rm OutTE}(X)$, where $\hat \theta$ denotes the estimate of a measure $\theta$, obtained from the data. In fact, the estimates $\hat H(Y(t)/Y_-)$ and $\hat H(Y(t) / X_-,Y_-)$ of the conditional entropies in Eq.(\ref{ce}) tend to underestimate the true values when these values are nonzero and the sample size is small, due to insufficient observation of events~(Appendix B). From the fact that  $\hat H(Y(t)/Y_-) \geq \widehat H(Y(t) / X_-,Y_-)$~(Appendix A), we see that if the underestimation is so severe that $\hat H (Y(t)/Y_-) \simeq 0$, then  $\hat H (Y(t) / X_-,Y_-) \simeq 0$,   leading to $\hat T_{X \to Y} = \hat H (Y(t) / Y_-)  - \hat H (Y(t) / X_-,Y_-) \simeq 0$. On the other hand, we encounter an overestimation problem when $T_{X \to Y}=H(Y/Y_-)- H(Y/X_-,Y_-)=0$. In this case,  due to estimation errors, $\hat H(Y/X_-,Y_-) \ne H(Y/X_-,Y_-)$ and $\hat H(Y/Y_-) \ne H(Y/Y_-)$, and there is no guarantee that $\hat H(Y/X_-,Y_-)$ and $\hat H(Y/Y_-)$ will cancel each other out. As a result, we have  $\hat T_{X \to Y} = \hat H(Y/Y_-)- \hat H(Y/X_-,Y_-) \geq 0$, leading to an overestimation. 

\begin{figure}
\includegraphics[width=\textwidth]{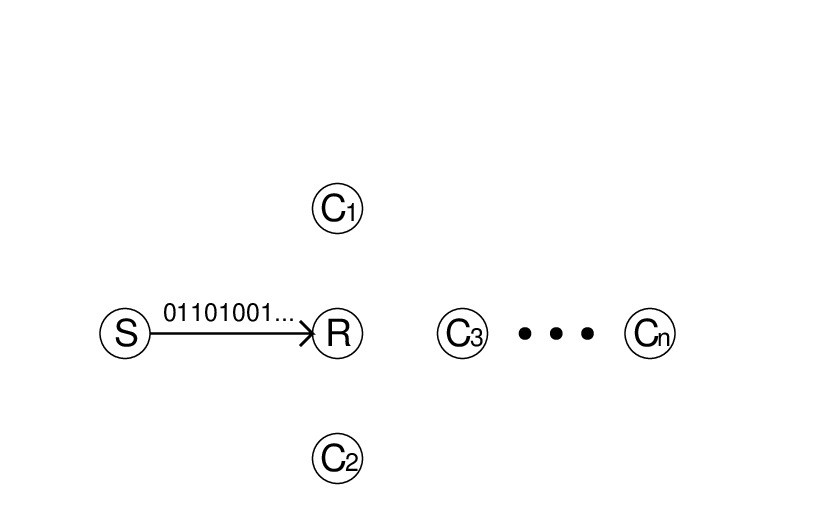}
\caption{The SR model with a sender node S and a receiver node R. The nodes ${\rm C}_1, \cdots {\rm C}_n$ represent confounding nodes that are independent of the dynamics between S and R.}
\label{model1} 
\end{figure}

To illustrate these issues, consider a simple system consisting of a sender node S and a recipient node R, referred to as the SR model (Fig.~\ref{model1}). The state of each node, denoted as $S(t)$ and $R(t)$, is either $0$ or $1$. The state $S(t)$ is completely random, chosen between 0 and 1 with equal probability. $R(0)$ is also random, but for $t >0$, $R(t)$ is fully determined by $S(t-1)$ and $R(t-1)$ according to the rule $R(t) = S(t-1) \oplus R(t-1)$, where $\oplus$ denotes the exclusive OR operation. Without knowledge of the sender’s state, the dynamics of the recipient node appear entirely random, leading to $H (R(t)/R_-)=1$\ bit. Furthermore, since $R(t)$ is fully determined by $S(t-1)$ and $R(t-1)$, we have $H (R(t)/R_-,S_-)=0$, resulting in $T_{S\to R} = 1$\ bit. In contrast, since $H(S(t)/S_-)= H(S(t)/R_-,S_-) = 1$ bit, it follows that $T_{R\to S} = 0$. Note that this system exhibits Markovian dynamics, as the entire process depends only on the previous states of S and R.

The underestimation problem can be demonstrated by a small example of four observed state transitions $(S, R)$:
\begin{eqnarray}
    (0,0) &\to& (0,0),\quad (0,1) \to (1,1), \quad (1,0) \to (1,1), \quad (1,1) \to (1,0). \label{tran1}
\end{eqnarray}
In this case, the empirical conditional probability distributions $\hat P(R(t)/R_-)$ and $\hat P(R(t)/S_-,R_-)$ coincide with the true conditional probability distributions $P(R(t)/R_-)$ and $P(R(t)/S_-,R_-)$, respectively, so $\hat T_{S\to R}= T_{S\to R}=1$\ bit. 

Next, suppose we add some nodes labeled ${\rm C}_1, {\rm C}_2, \cdots {\rm C}_n$  whose states take random values of $0$ or $1$ and are entirely disconnected from the rest of the system.  I will refer to these as ``confounding nodes"~(see Fig.~\ref{model1}). The true values of ${\rm OutTE}(S)$ and ${\rm OutTE}(R)$ remain unaffected, as the dynamics of ${\rm C}_1, {\rm C}_2, \cdots {\rm C}_n$ are independent of S and R. However, they can severely impact the {\it estimated} values of ${\rm OutTE}(S)$ and ${\rm OutTE}(R)$. Let us assume just one confounding node C is added to the example in Eq.(\ref{tran1}) so that the four observed transitions of the states $(S, R, C)$ become:
\begin{eqnarray}
    (0,0,1) &\to& (0,0,1),\quad (0,1,0) \to (1,1,1), \quad (1,0,0) \to (1,1,0), \quad (1,1,1) \to (1,0,0) \label{tran2}
\end{eqnarray}
Now the dynamics of $(R,C)$ {\it estimated} by the data above is completely deterministic. Specifically, the only four observed transitions for $(R, C)$ are
\begin{eqnarray}
    (0,1) &\to& (0,1),\quad (1,0) \to (1,1), \quad (0,0) \to (1,0), \quad (1,1) \to (0,0), 
\end{eqnarray}
which are all unique. This leads to $\hat H(R(t),C(t))/R_-,C_-)= \hat H(R(t),C(t))/R_-,C_-,S_-)= 0$, resulting in $\widehat {\rm OutTE}(S) = \hat T_{S\to (R,C)} = 0$. This represents a drastic underestimation compared to the true value of ${\rm OutTE}(S) = 1$. Such an artifact occurs when the number of variables is comparable to or larger than the number of samples. In such situations,  most observed transitions become rare events, typically appearing only once in the data, leading to the underestimation of the values of conditional entropy~(Appendix B). As for the  overestimation, specific examples will be discussed below.

\begin{figure}
\includegraphics[width=\textwidth]{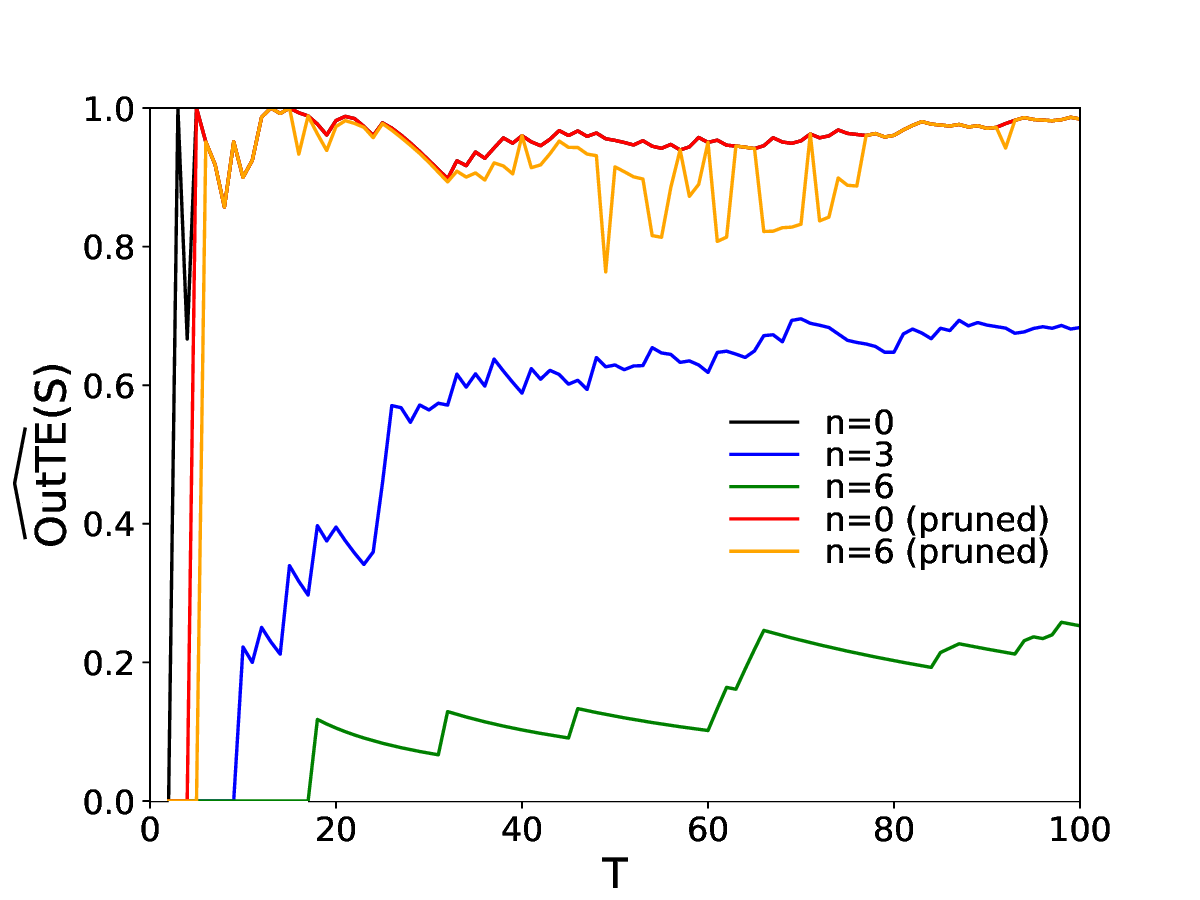}
\caption{$\widehat {\rm OutTE}(S)$ in the SR model as a function of $T$, without pruning (black, blue, and green lines for $n=0, 3, 6$, respectively), and with pruning (red and orange lines for $n=0$ and $6$, respectively).}
\label{outs} 
\end{figure}

\begin{figure}
\includegraphics[width=\textwidth]{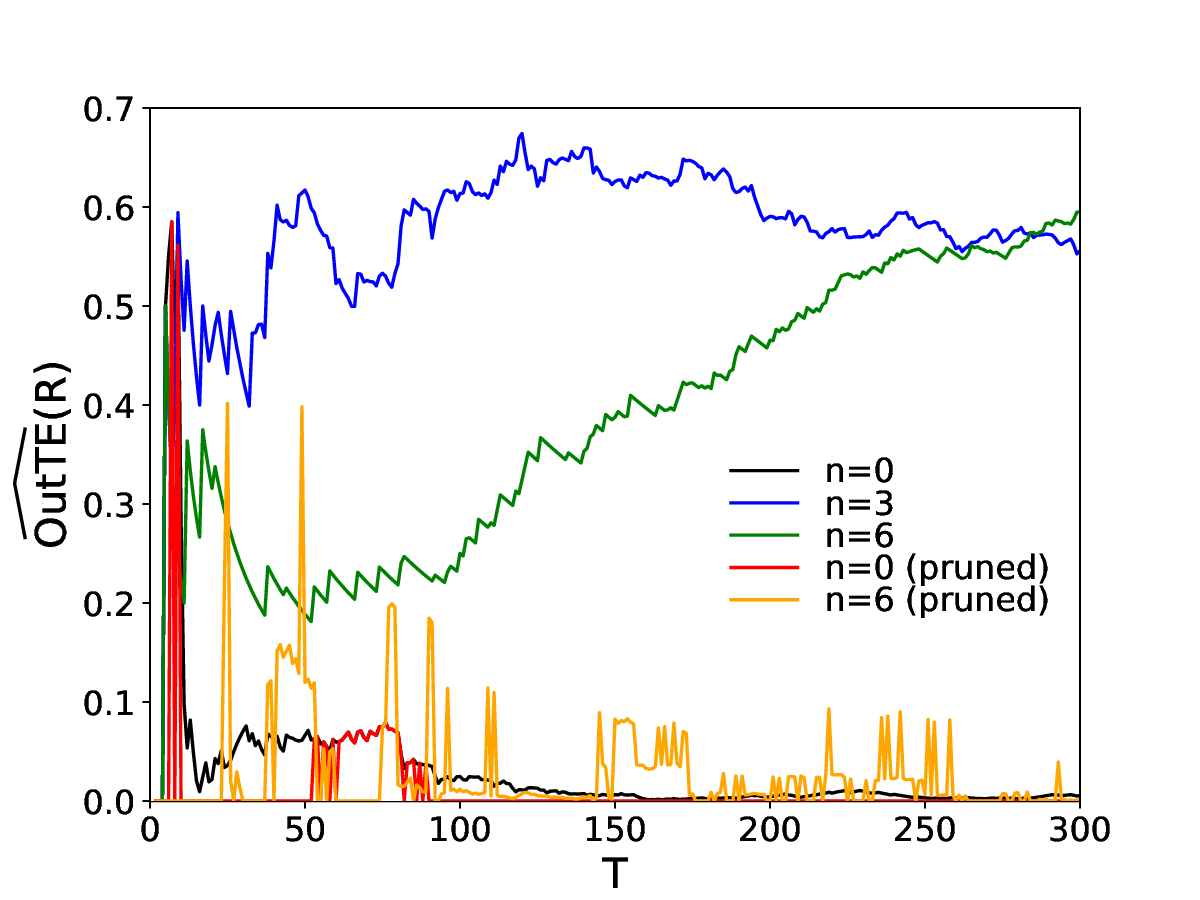}
\caption{$\widehat {\rm OutTE}(R)$ in the SR model as a function of $T$, without pruning (black, blue, and green lines for $n=0, 3, 6$, respectively), and with pruning (red and orange lines for $n=0$ and $6$, respectively).}
\label{outr} 
\end{figure}

\begin{figure}
\includegraphics[width=\textwidth]{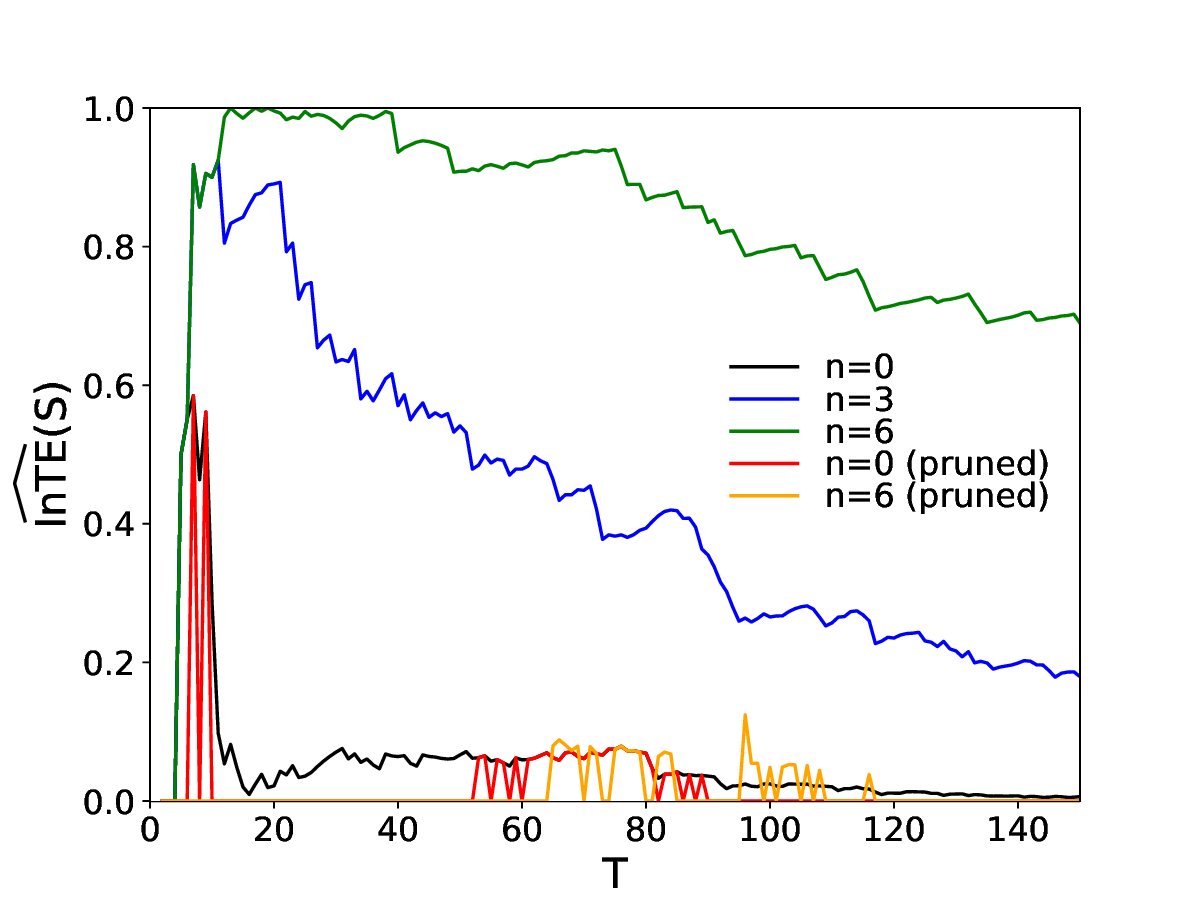}
\caption{$\widehat {\rm InTE}(S)$ in the SR model as a function of $T$, without pruning (black, blue, and green lines for $n=0, 3, 6$, respectively), and with pruning (red and orange lines for $n=0$ and $6$, respectively).}
\label{ins} 
\end{figure}

\begin{figure}
\includegraphics[width=\textwidth]{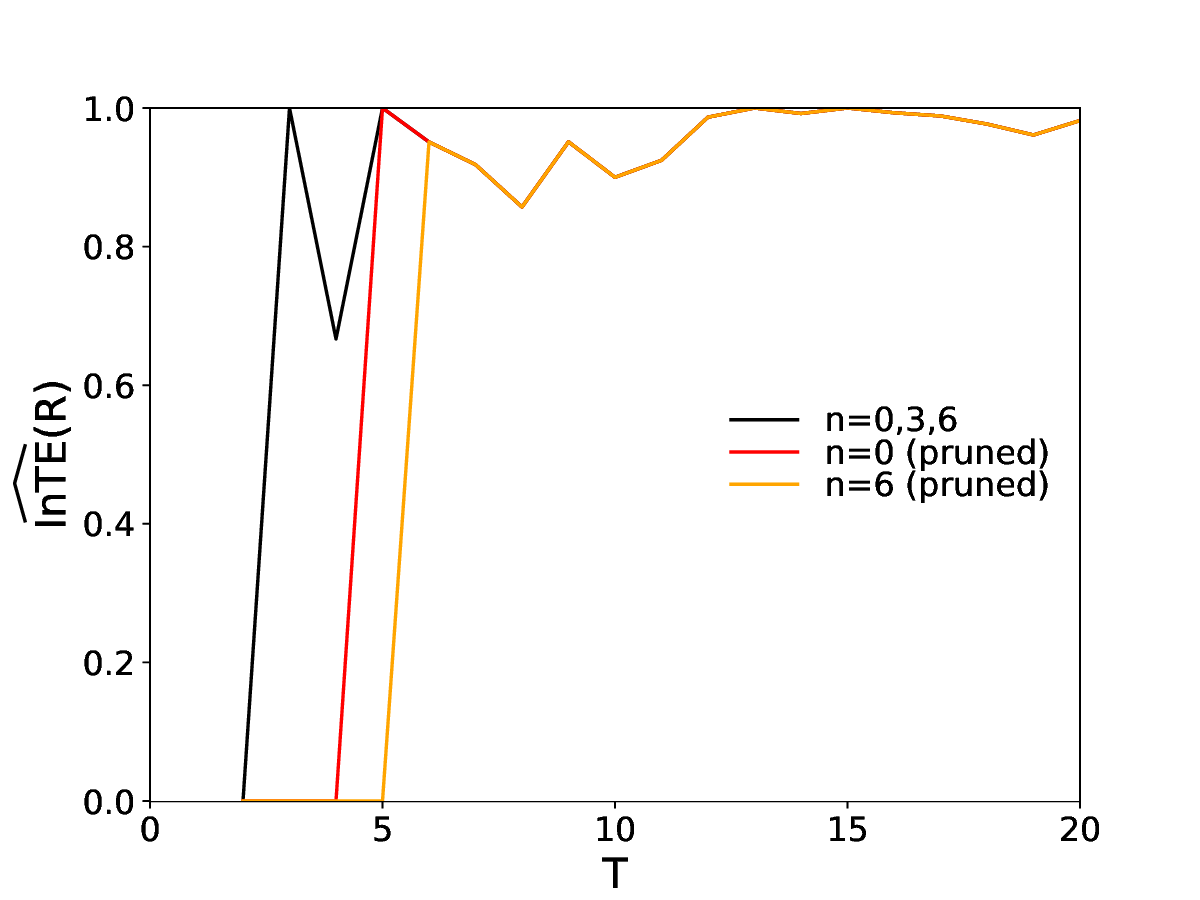}
\caption{$\widehat {\rm InTE}(R)$ in the SR model as a function of $T$, without pruning (black line for $n=0, 3, 6$), and with pruning (red and orange lines for $n=0$ and $6$, respectively).}
\label{inr} 
\end{figure}

A synthetic time series of length 300 were generated using the probability distribution of the SR model, and the estimators $\widehat {\rm OutTE}$ and $\widehat {\rm InTE}$ were computed from the empirical distribution using partial series of length $T$. I assumed Markovian dynamics from the start, so the number of observed transitions is $T-1$. $\widehat {\rm OutTE}$ and $\widehat {\rm InTE}$ were computed using the JIDT toolkit~\cite{jidt}.  The graphs of $\widehat {\rm OutTE}(S)$, $\widehat {\rm OutTE}(R)$,  $\widehat {\rm InTE}(S)$ and $\widehat {\rm InTE} (R)$ are shown as functions of $T$ in Figs.~\ref{outs}, \ref{outr}, \ref{ins}, and \ref{inr}, respectively, for different numbers of confounding nodes: $n= 0$ (black line), $n=3$ (blue line), and $n=6$ (green line). We observe that $\widehat {\rm OutTE}(S)$ is lower than the true value of ${\rm OutTE}(S)=1$\ bit for small sample sizes, but begins to converge to the true value around $T \simeq 10$ for $n=0$ (Fig.~\ref{outs}). As the confounding nodes are added, the underestimation becomes more severe for a given  $T$, and convergence becomes slower. In the case of  ${\rm OutTE}(R)$, whose true value is zero, we now encounter an overestimation problem in the presence of confounding nodes (Fig.~\ref{outr}).  We also observe that the estimation error increases with $T$ over the range shown in the figure, and that the estimation with $n=6$ is better for small $T$, but becomes worse for $T \gtrsim 300$. This occurs because, for small $T$ and large $n$, $\hat H(S(t), C_1(t), \cdots, C_n(t))/(S(t-1), C_1(t-1), \cdots, C_n(t-1))\simeq 0$ and $\hat H(S(t), C_1(t), \cdots, C_n(t))/(R(t-1), S(t-1), C_1(t-1), \cdots, C_n(t-1)) \simeq 0$, due to the underestimation problem, resulting in $\widehat {\rm OutTE}(R) \simeq 0$. Thus, the small value of $\widehat {\rm OutTE}(R)$ for small $T$ and large $n$ is an artifact of the limited sample size rather than an accurate estimation. 

For ${\rm InTE}(S)$, whose true value is  zero, we also encounter an overestimation problem, but the artifact observed for $\widehat {\rm OutTE}(R)$ does not appear here because the target is univariate. The overestimation becomes progressively worse as the confounding nodes are added (Fig.~\ref{ins}). Finally, the estimate of  ${\rm InTE}(R)$, whose true value is one bit, remains unaffected by the presence of the confounding nodes, as shown by the black line in Fig.~\ref{inr}. Note that according to the definition of the transfer entropy in Eq.~(\ref{te2}), the estimate $\widehat {\rm InTE}(R)$ of ${\rm InTE}(R)$ can be written as
\begin{equation}
    \widehat {\rm InTE}(R) \equiv \hat T_{{\rm rest} \to R} =  \hat T_{(S, C_1, \cdots, C_N) \to R} = \hat H(R(t)/R_-)- \hat H(R(t)/R_-, S_-, {C_1}_-, \cdots, {C_N}_-).\label{te4}
\end{equation}
Since the first term $\hat H(R(t)/R_-)$ in Eq.~(\ref{te4}) does not depend on the confounding nodes, the only possible dependence on the confounding nodes originates from the second term $\hat H(R(t)/R_-, S_-, {C_1}_-, \cdots, {C_N}_-)$. However, in this model, the values of $R(t-1)$ and $S(t-1)$ uniquely determine the value of $R(t)$, which also holds true for the observed transitions. The addition of confounding nodes does not affect this determinism. Therefore, $\hat H(R(t)/R_-, S_-, {C_1}_-, \cdots, {C_N}_-)=0$ regardless of the number of confounding nodes, leading to $\widehat {\rm InTE}(R)  = \hat H(R(t)/R_-)$, which is independent of the confounding nodes. 

\begin{figure}
\includegraphics[width=\textwidth]{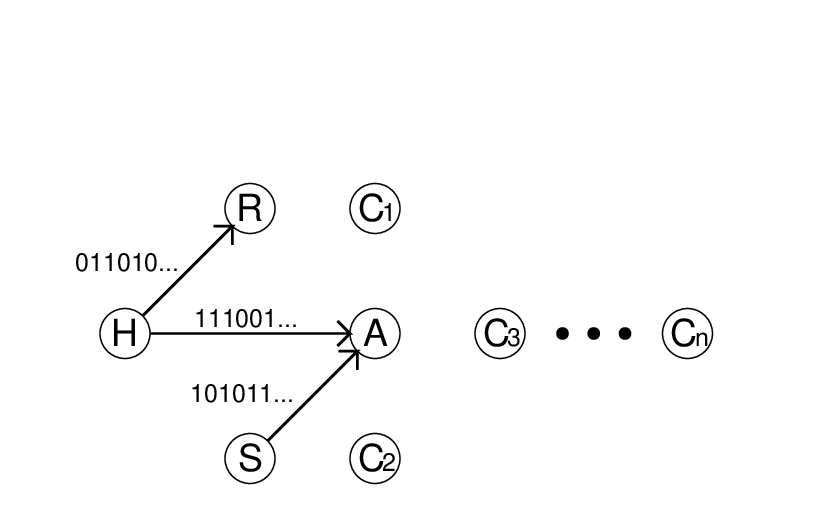}
\caption{The HSAR model with a hub node H, a sender node S, an anti-hub node A,  and a receiver node R. The nodes ${\rm C}_1, \cdots {\rm C}_n$ represent confounding nodes that are independent of the dynamics among H, S, A, and R.}
\label{model2} 
\end{figure}

\begin{figure}
\includegraphics[width=\textwidth]{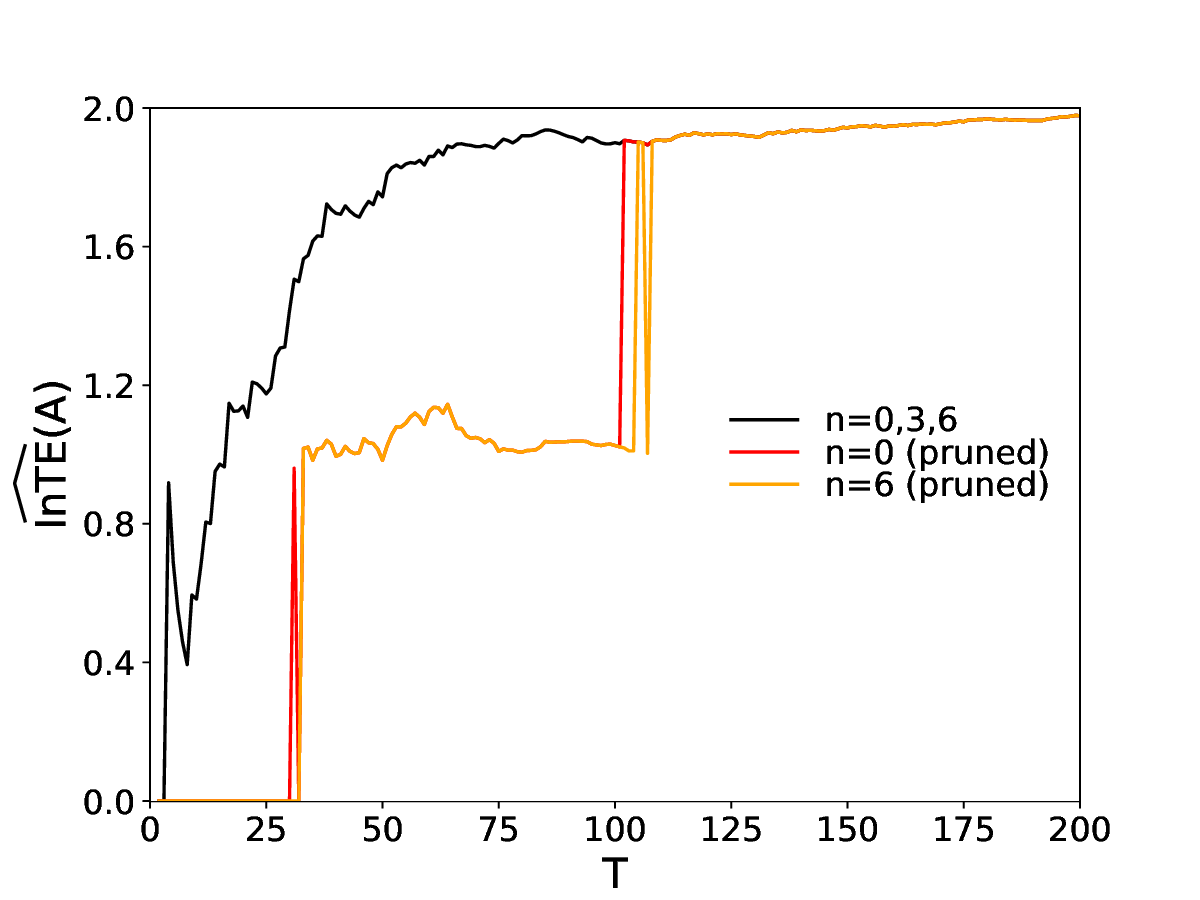}
\caption{$\widehat {\rm InTE}(A)$ in the HSAR model as a function of $T$, without pruning  (black line for $n=0, 3, 6$), and with pruning (red and orange lines for $n=0$ and $6$, respectively).}
\label{inhaa} 
\end{figure}

\begin{figure}
\includegraphics[width=\textwidth]{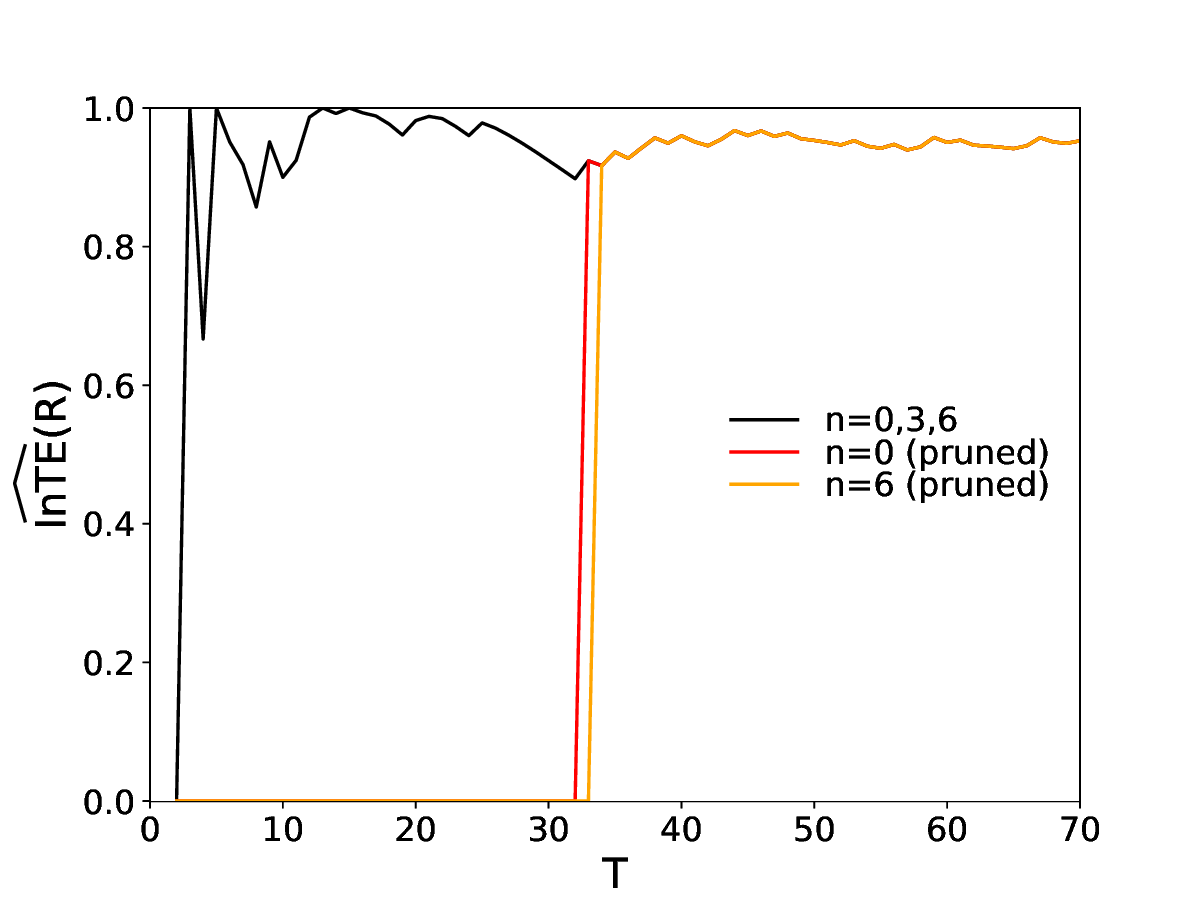}
\caption{$\widehat {\rm InTE}(R)$ in the HSAR model as a function of $T$, without pruning (black line for $n=0, 3, 6$), and with pruning (red and orange lines for $n=0$ and $6$, respectively).}
\label{inhar} 
\end{figure}

I conducted a similar analysis on a system containing a hub H, a sender node S, an anti-hub A, and a receiver node R, as shown in Fig.~\ref{model2}. I will refer to this system as the HSAR model. In this model, the hub node H sends two independent bits of information to two nodes A and R, and S sends one bit of information to A. As a result, A receives one bit of information from both H and S (total two bits), while R receives one bit of information from H alone. The estimates $\widehat {\rm OutTE}(H)$, $\widehat {\rm OutTE}(S)$, $\widehat {\rm OutTE}(A)$, $\widehat {\rm OutTE}(R)$, $\widehat {\rm InTE}(H)$,  $\widehat {\rm InTE}(S)$,  $\widehat {\rm InTE}(A)$, and $\widehat {\rm InTE}(R)$, are shown in Figs.\ref{outhah} through \ref{inhar} for different numbers of confounding nodes ($n=0,3,6$), represented by black, blue, and green lines, respectively. 

The results are qualitatively similar to those of the SR model.  Although  $\widehat {\rm OutTE}(H)$ and $\widehat {\rm OutTE}(S)$ converge towards their true values  (2 bits for H and 1 bit for S) as $T$ increases, the convergence slow down for larger $n$.  For ${\rm OutTE}(A)$ and ${\rm OutTE}(R)$, the estimates are larger than their true values (which are zero). However, this overestimation is mitigated for large $n$ and small $T$, due to the same artifact as in the case of  $\widehat {\rm OutTE}(R)$ in the SR model. The overestimation of ${\rm InTE}(H)$ and ${\rm InTE}(S)$, whose true values are zero, worsens with increasing $n$, much like the overestimation of ${\rm InTE}(S)$ in the SR model. Finally, $\widehat {\rm InTE}(A)$ and $\widehat {\rm InTE}(R)$ remain unaffected by the presence of confounding nodes for the same reason as in  $\widehat {\rm InTE}(R)$ in the SR model.

\section{The Method for accurate estimation of Outgoing and Incoming transfer entropy}
In the previous examples, the estimation errors were due to the proliferation of variables caused by confounding nodes. If we had eliminated unrelated nodes beforehand, the estimation error would have been reduced. Therefore,  instead of estimating OutTE from a node to all  other nodes, and InTE to the node from all others, we first take a pruning step where only nodes causally related the node of interest are selected. That is, we construct a directed binary network where an edge $X \to Y$ exists if and only if $\widehat T_{(X \to Y)/Z}$ is statistically significant. The construction of such a network using the exact computation of $\widehat T_{(X \to Y)/Z}$, along with rigorous statistical tests, is computationally costly, and  various approximation schemes have been proposed to estimate $T_{(X \to Y)/Z}$ and construct the causal network~\cite{vla10, faes11,   mont14, sun15, runge12b, lizi19}. Here, I used the method developed in ref.~\cite{lizi19}, where for each target node, the set of source nodes with statistically meaningful causal influence is constructed step by step. This method has been implemented as a publicly available Python package~\cite{woll19} and has been used for brain record data consisting of 100 nodes and 10,000 samples~\cite{lizi19}. 

\begin{figure}
\includegraphics[width=\textwidth]{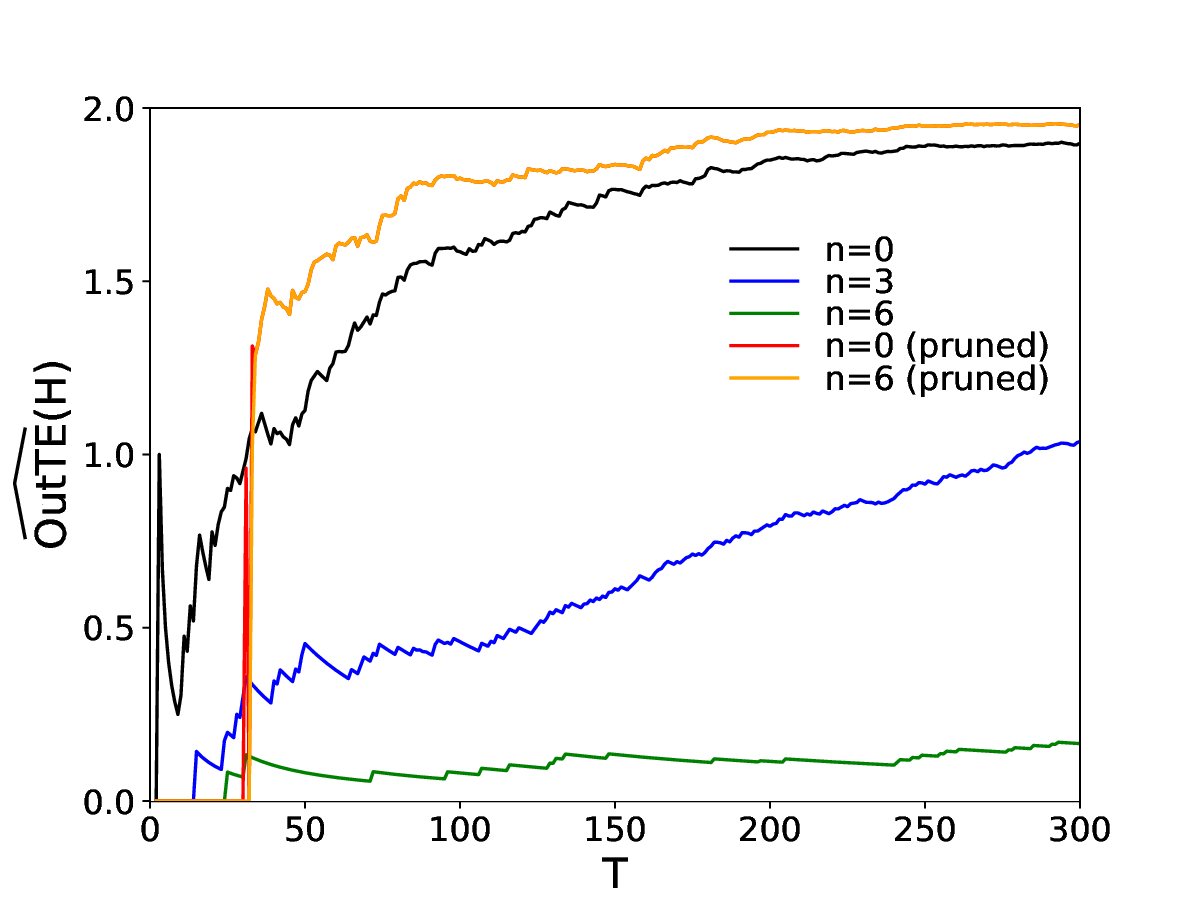}
\caption{$\widehat {\rm OutTE}(H)$ in the HSAR model as a function of $T$, for $n=0, 3, 6$ without pruning (black, blue, and green lines for $n=0, 3, 6$, respectively) and with pruning (red and orange lines for $n=0$ and $6$, respectively).}
\label{outhah} 
\end{figure}

\begin{figure}
\includegraphics[width=\textwidth]{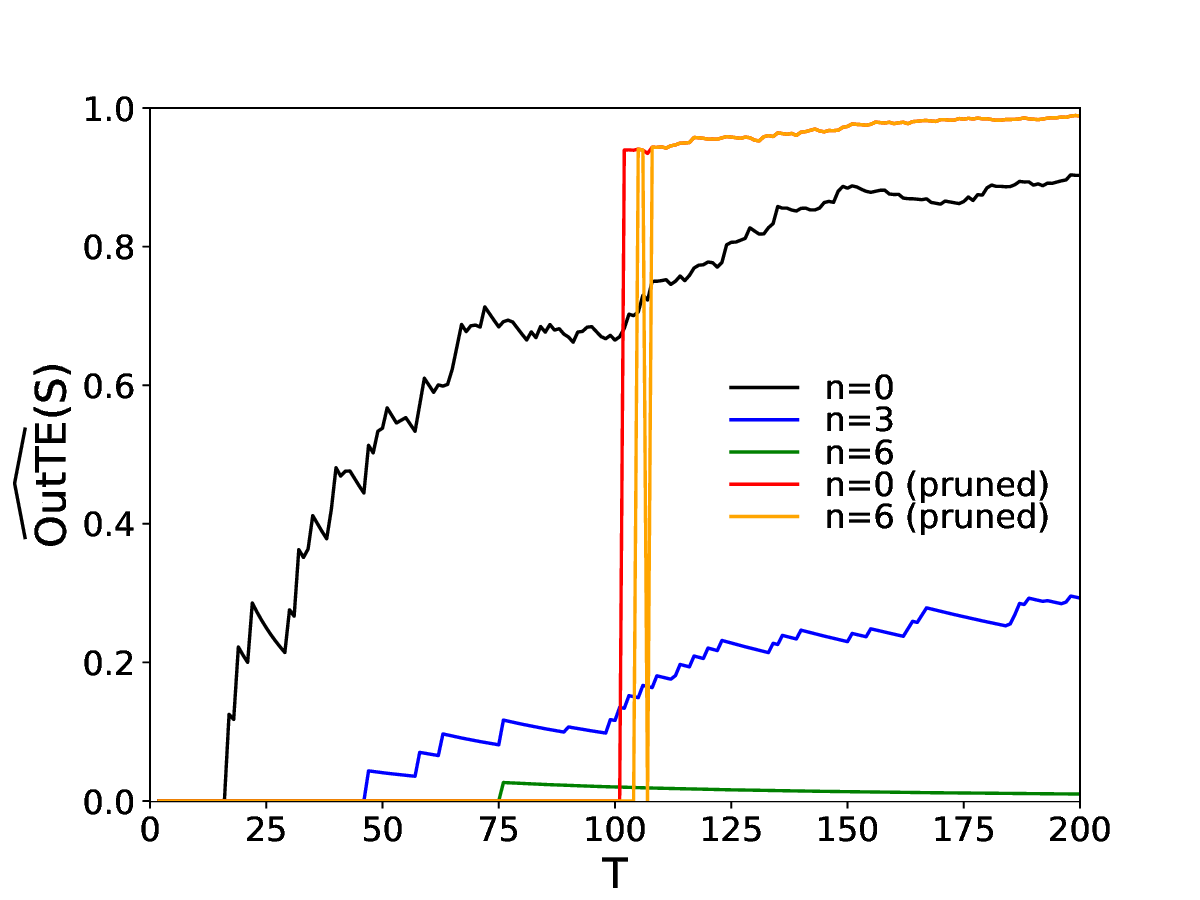}
\caption{$\widehat {\rm OutTE}(S)$ in the HSAR model as a function of $T$, without pruning  (black, blue, and green lines for $n=0, 3, 6$, respectively), and with pruning (red and orange lines for $n=0$ and $6$, respectively).}
\label{outhas} 
\end{figure}

\begin{figure}
\includegraphics[width=\textwidth]{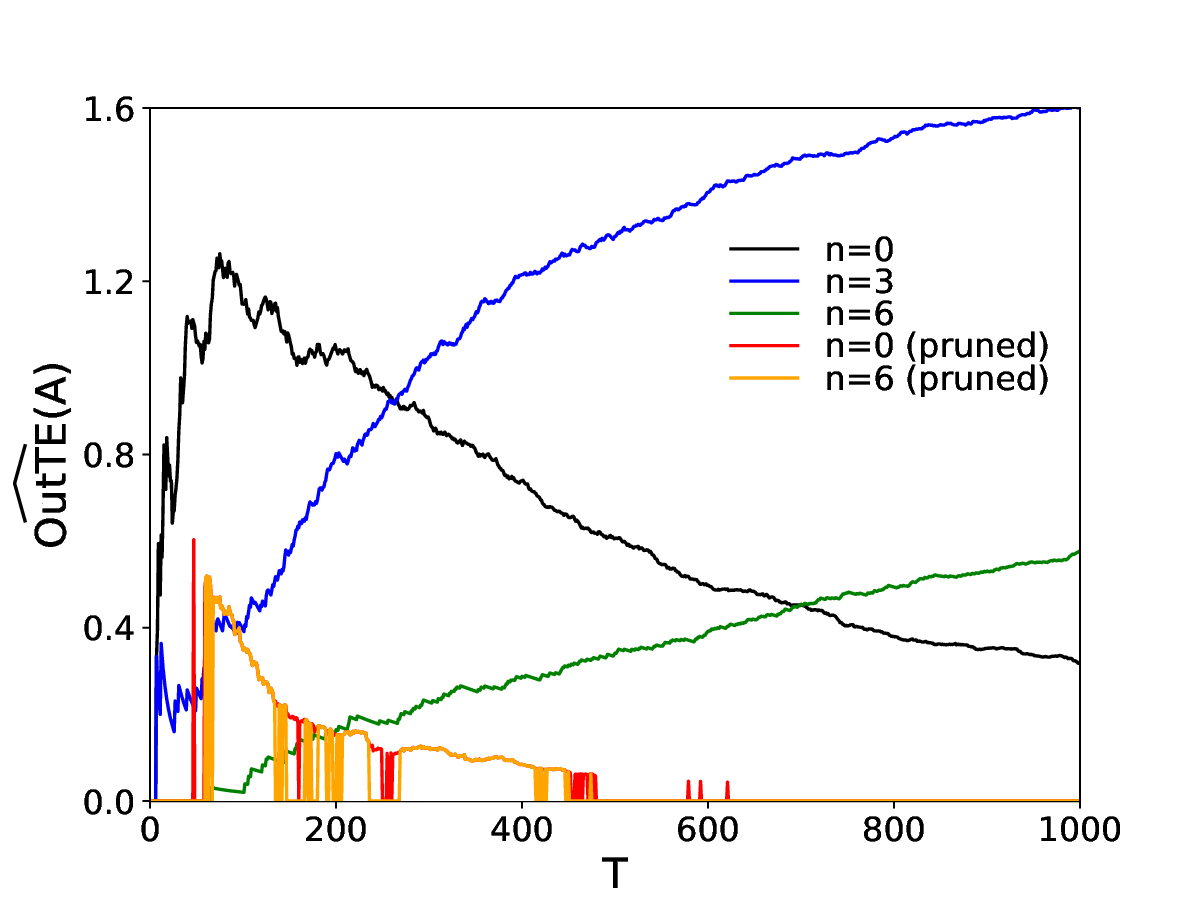}
\caption{$\widehat {\rm OutTE}(A)$ in the HSAR model as a function of $T$, without pruning  (black, blue, and green lines for $n=0, 3, 6$, respectively), and with pruning (red and orange lines for $n=0$ and $6$, respectively).}
\label{outhaa} 
\end{figure}

\begin{figure}
\includegraphics[width=\textwidth]{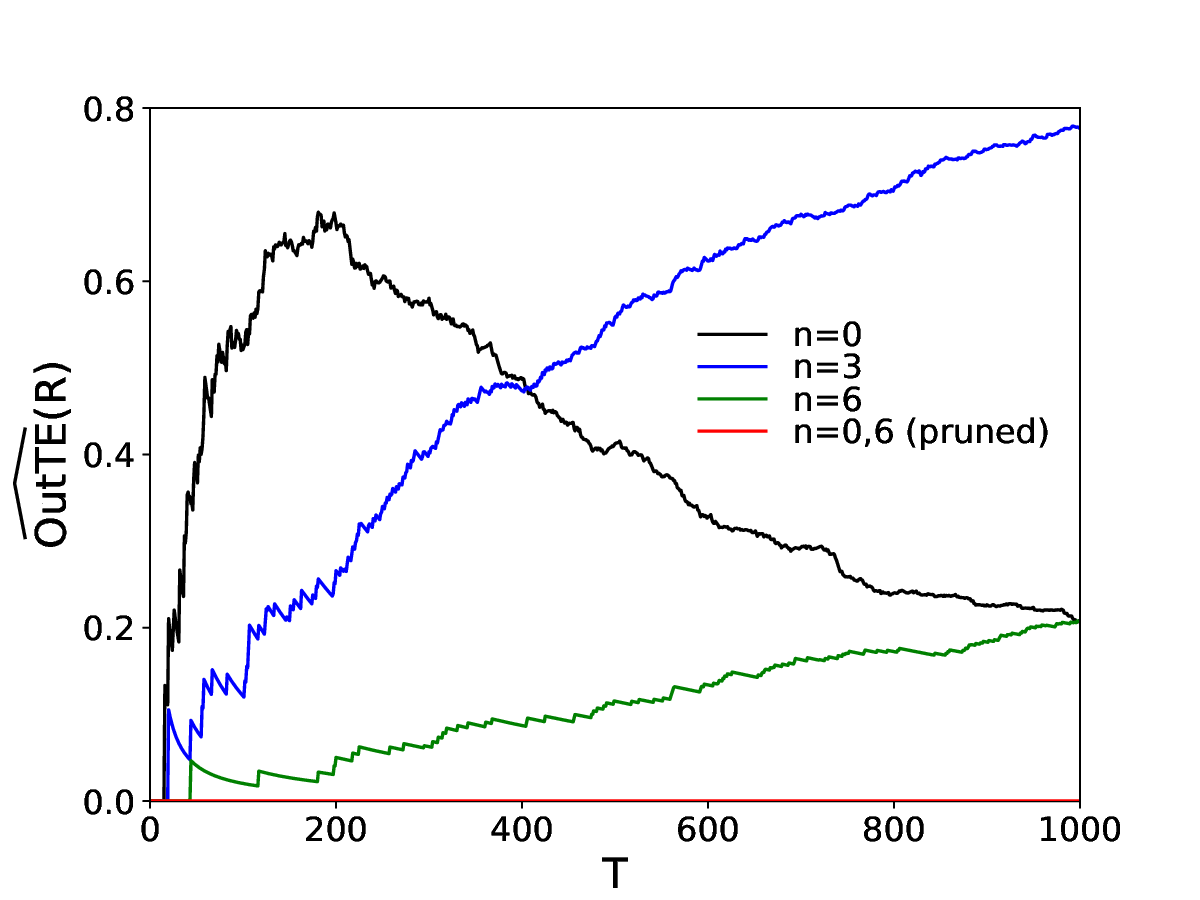}
\caption{$\widehat {\rm OutTE}(R)$ in the HSAR model as a function of $T$, without pruning  (black, blue, and green lines for $n=0, 3, 6$, respectively), and with pruning (red line for $n=0$ and $6$).}
\label{outhar} 
\end{figure}

\begin{figure}
\includegraphics[width=\textwidth]{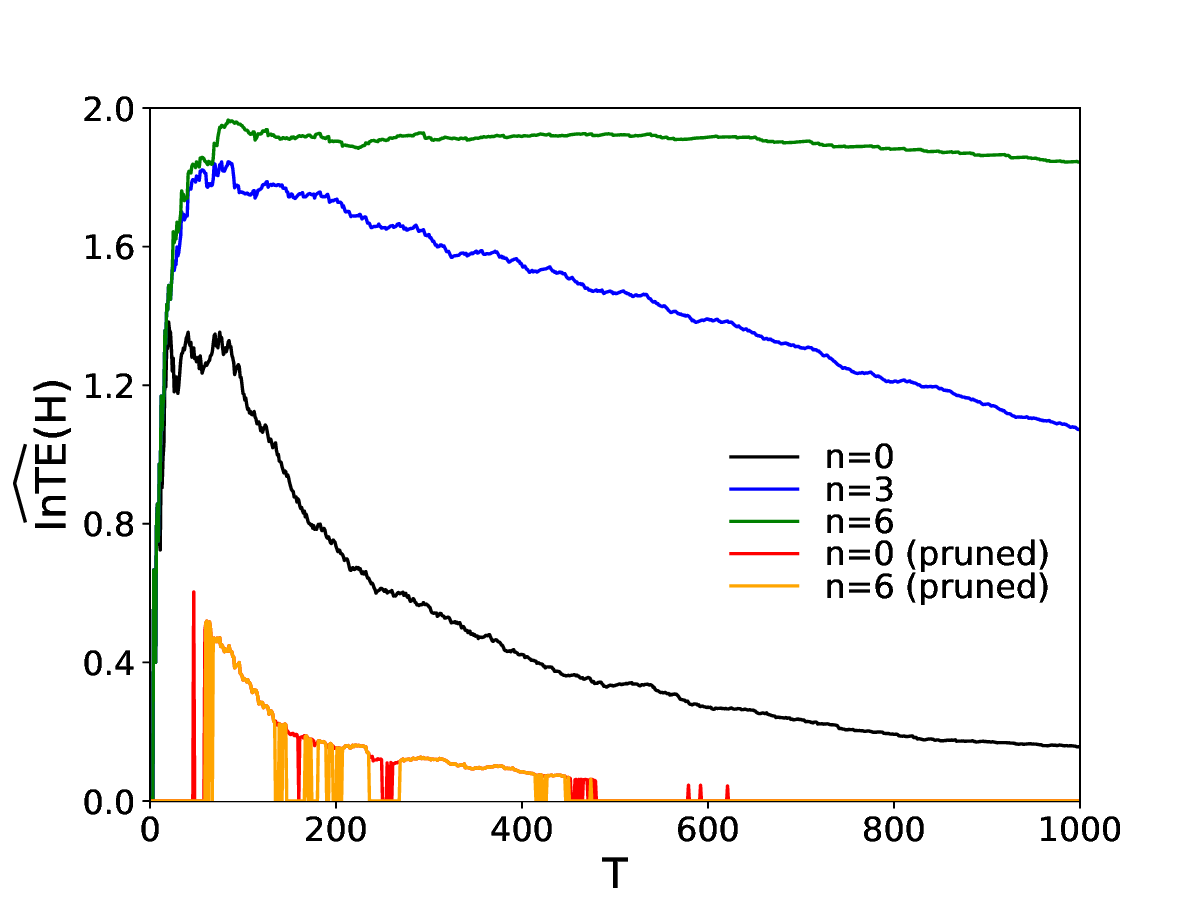}
\caption{$\widehat {\rm InTE}(H)$ in the HSAR model as a function of $T$, without pruning  (black, blue, and green lines for $n=0, 3, 6$, respectively), and with pruning (red and orange lines for $n=0$ and $6$, respectively).}
\label{inhah} 
\end{figure}

\begin{figure}
\includegraphics[width=\textwidth]{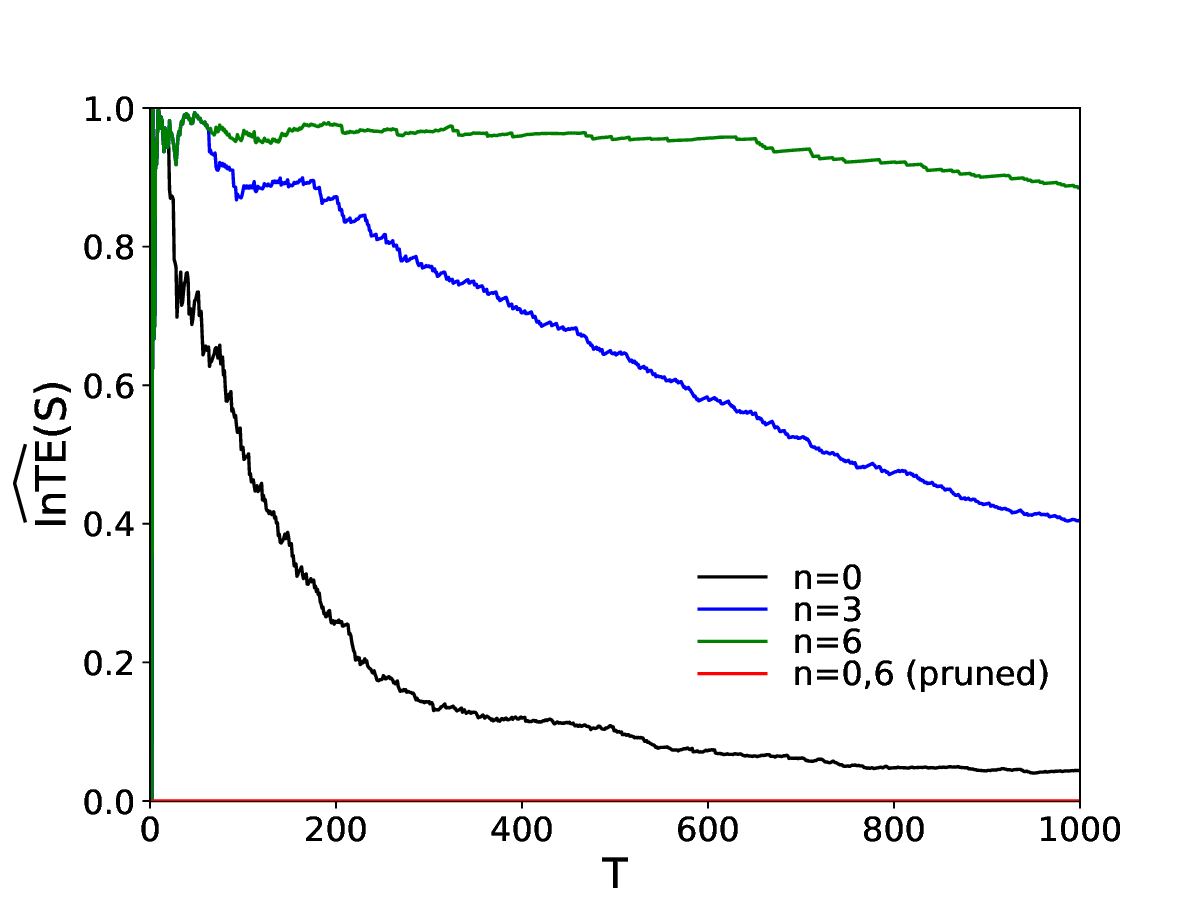}
\caption{$\widehat {\rm InTE}(S)$ in the HSAR model as a function of $T$, without pruning  (black, blue, and green lines for $n=0, 3, 6$, respectively), and with pruning (red line for $n=0$ and $6$).}
\label{inhas} 
\end{figure}

By computing OutTE and InTE only between the node of interest and those connected by edges in the binary network, estimation errors are significantly reduced, as shown in Figs.~\ref{outs}-\ref{inr} for the SR model and Figs.~\ref{outhah}-\ref{inhar} for the HSAR model 
for $n=0$ (red lines) and $n=6$ (orange lines). We find that the estimation error with pruning does not significantly increase even as the number of confounding nodes rises from 0 to 6. 

In the SR model, the pruned estimate $\widehat {\rm OutTE}(S)$ quickly converges to the true value of 1 bit around $T \simeq 10$~(Fig.~\ref{outs}) regardless of $n$, overcoming the underestimation problem. Pruning also reduces the overestimation problem of $\widehat {\rm OutTE}(R)$, even for $n=6$,  where $\widehat {\rm OutTE}(R) < 0.1$ for $T \geq 112$~(Fig.~\ref{outr}). The same applies to $\widehat {\rm InTE}(S)$,  where $\widehat {\rm InTE}(S) < 0.1$ for $T \geq 97$ even for $n=6$~(Fig.~\ref{ins}). In case of  $\widehat {\rm InTE}(R)$ where confounding nodes are not problematic, pruning is not useful: in fact, it increases estimation error for very small sample sizes. However, this error quickly diminishes once $T$ reaches 6~(Fig.~\ref{inr}), so the damage is minimal.

In the HSAR model, the pruned estimate $\widehat {\rm OutTE}(H)$  is zero up to $T \simeq 30$ because the network construction algorithm could not detect the outgoing edges from the node $H$. However, the estimate quickly converges to the true value of 2 bits for $T \gtrsim 30$ regardless of $n$~(Fig.~\ref{outhah}). In this range, the pruned estimate of ${\rm OutTE}(H)$ outperforms the unpruned estimate, even for $n=0$, as node S, which is unrelated to H in terms of ${\rm OutTE}(H)$, act as a confounding node for H. The same behavior is observed for $\widehat {\rm OutTE}(S)$, where the estimate with pruning converges to the true value of 1 bit around $T \simeq 100$.  Again, for $T \gtrsim 100$, pruning proves beneficial, even for $n=0$, because it eliminates the confounding effect from the node H~(Fig.~\ref{outhas}). Pruning also reduces the overestimation problem of $\widehat {\rm OutTE}(A)$~(Fig.~\ref{outhaa})  by removing the confounding effect from the node R as well as those from the nodes ${\rm C}_i\ (i=1, \cdots n)$. The effect of the pruning is even more drastic in the case of the estimate $\widehat {\rm OutTE}(R)$, since it makes the estimate coincide with the true value of zero, as the network algorithm does not detect any meaningful connection from the node R to any other node for any sample size~(Fig.~\ref{outhar}). The difference in the behavior of the network construction algorithm for ${\rm OutTE}(R)$ compared to ${\rm OutTE}(A)$ suggests that the presence of many incoming edges for A, which should not affect the true value of ${\rm OutTE}(A)$, somehow confuses the network construction algorithm at small sample sizes. This might result in the introduction of some false outgoing edges, leading to small but non-zero values of $\widehat {\rm OutTE}(A)$. 

Regarding InTE, pruning resolves the overestimation problem of ${\rm InTE}(H)$, where all of the S, R, and A nodes act as additional confounding nodes~(Fig.~\ref{inhah}).  For $\widehat {\rm InTE}(S)$, the pruned estimate coincides with the true value of 0 bits~(Fig.~\ref{inhas}). The difference between the effect of pruning on $\widehat {\rm InTE}(H)$ and $\widehat {\rm InTE}(S)$ is similar to that between $\widehat {\rm OutTE}(A)$ and $\widehat {\rm OutTE}(R)$. Pruning does not benefit $\widehat {\rm InTE}(A)$ or $\widehat {\rm InTE}(R)$, where confounding nodes are not problematic. However, introducing pruning does no harm for $T \gtrsim 100$ for $\widehat {\rm InTE}(A)$ and $T \gtrsim 30$ for $\widehat {\rm InTE}(R)$. The situation similar to that of $\widehat {\rm InTE}(R)$ in SR model. 

\section{Application to Microbiota data}
\begin{figure}
\includegraphics[width=\textwidth]{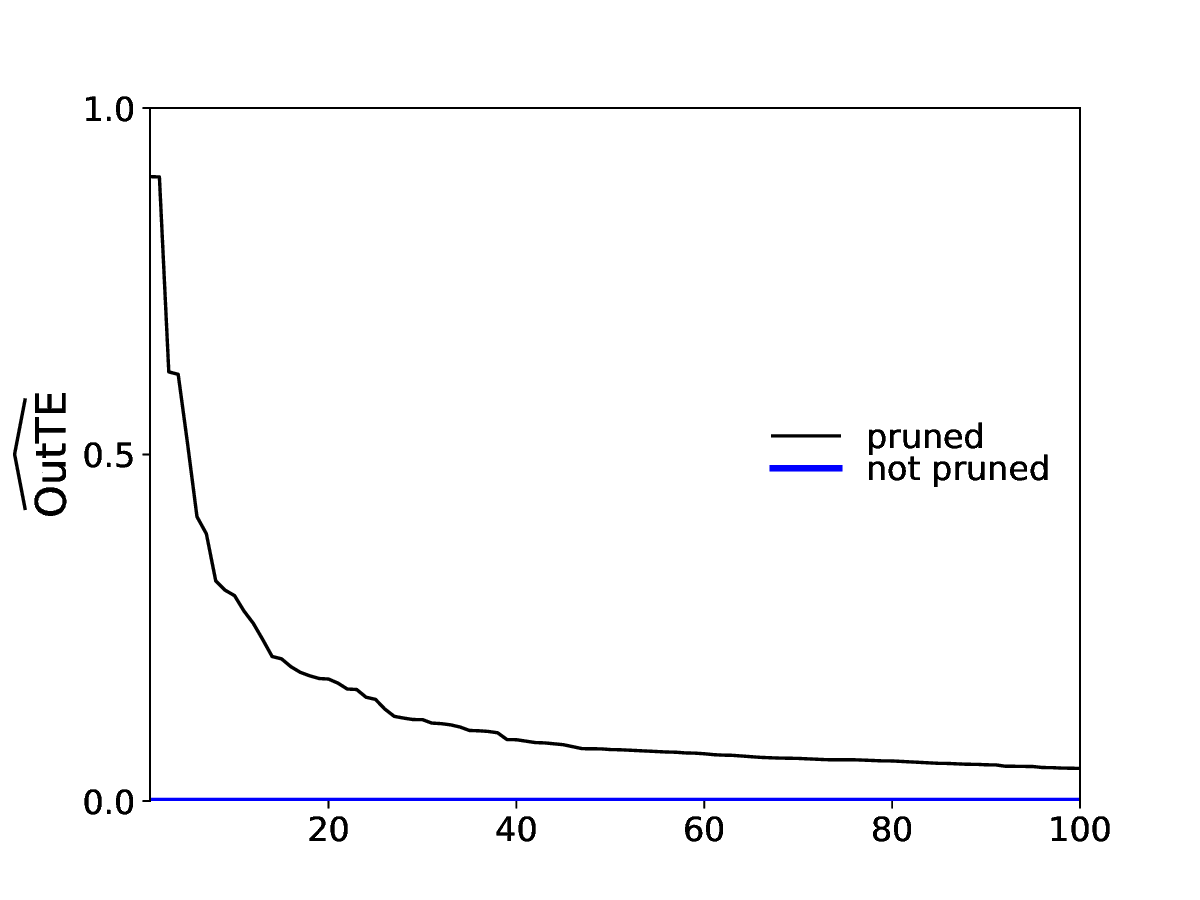}
\caption{$\widehat {\rm OutTE}$ of OTUs in the saliva microbiota data, obtained with pruning, sorted in descending order of their values (black line) shown for top 100 OTUs, compared with estimates obtained without pruning (blue line).}
\label{baout} 
\end{figure}

\begin{figure}
\includegraphics[width=\textwidth]{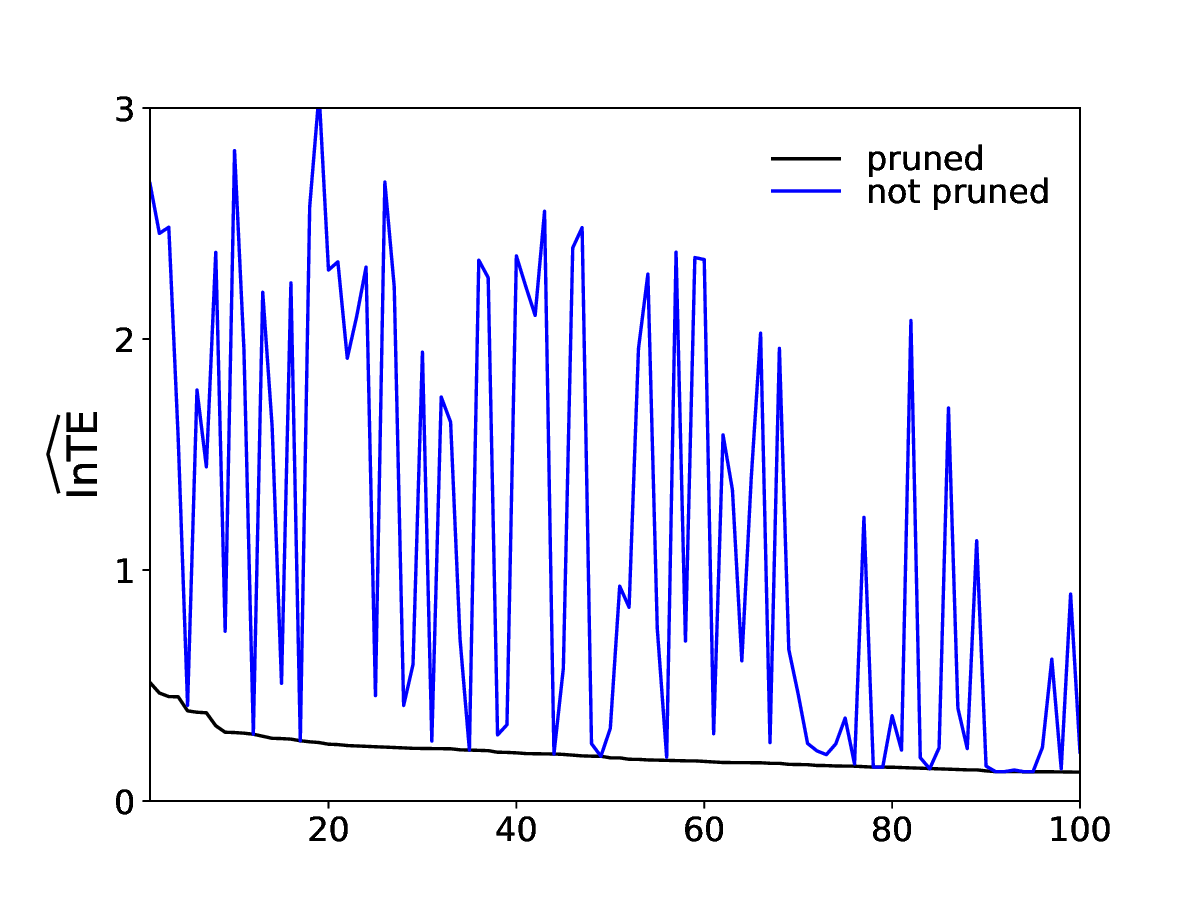}
\caption{$\widehat {\rm InTE}$ of OTUs in the saliva microbiota data, obtained with pruning, sorted in descending order of their values (black line) shown for top 100 OTUs, compared with estimates obtained without pruning (blue line).}
\label{bain} 
\end{figure}

\begin{figure}
\includegraphics[width=\textwidth]{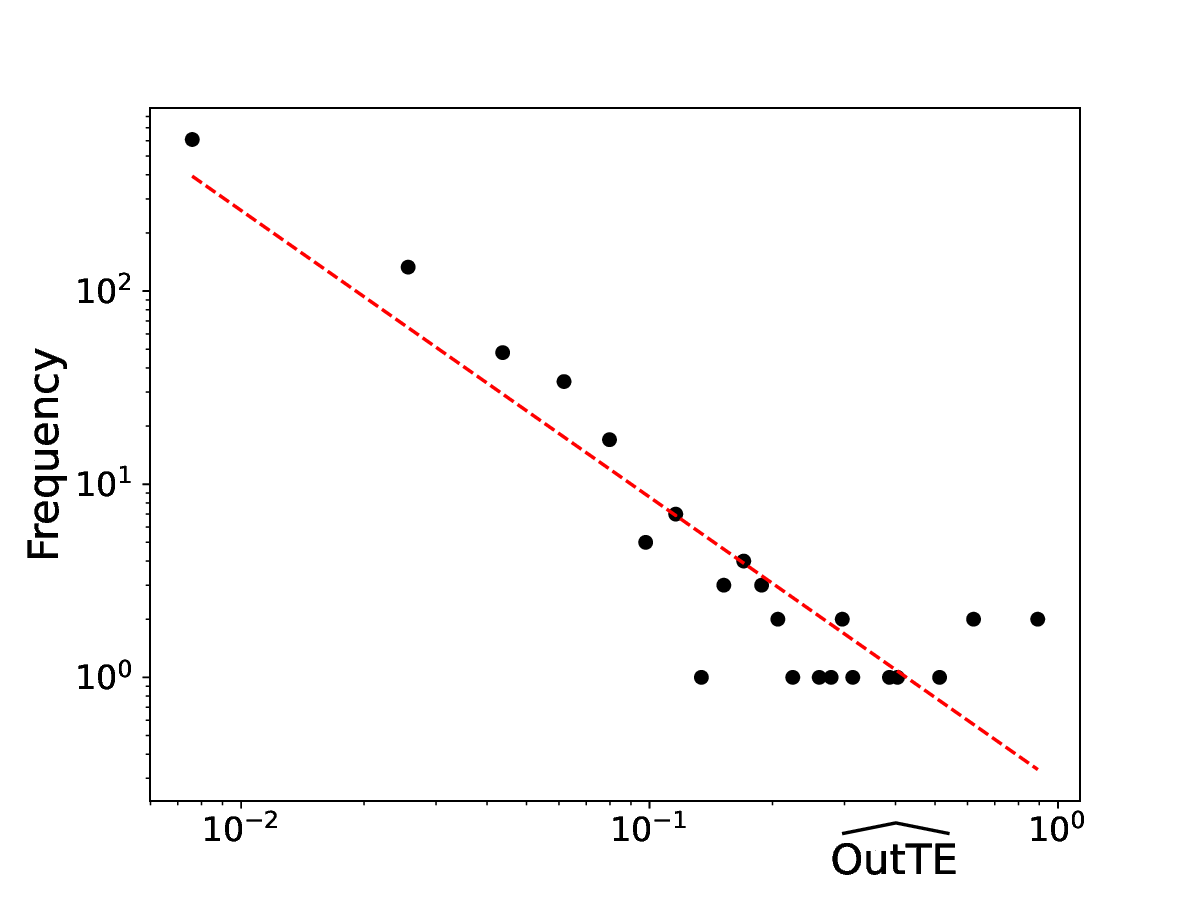}
\caption{Histrogram of $\widehat {\rm OutTE}$ (black dots) of OTUs in the saliva microbiota data, along with a power law fit (red dashed line)}
\label{histout} 
\end{figure}

\begin{figure}
\includegraphics[width=\textwidth]{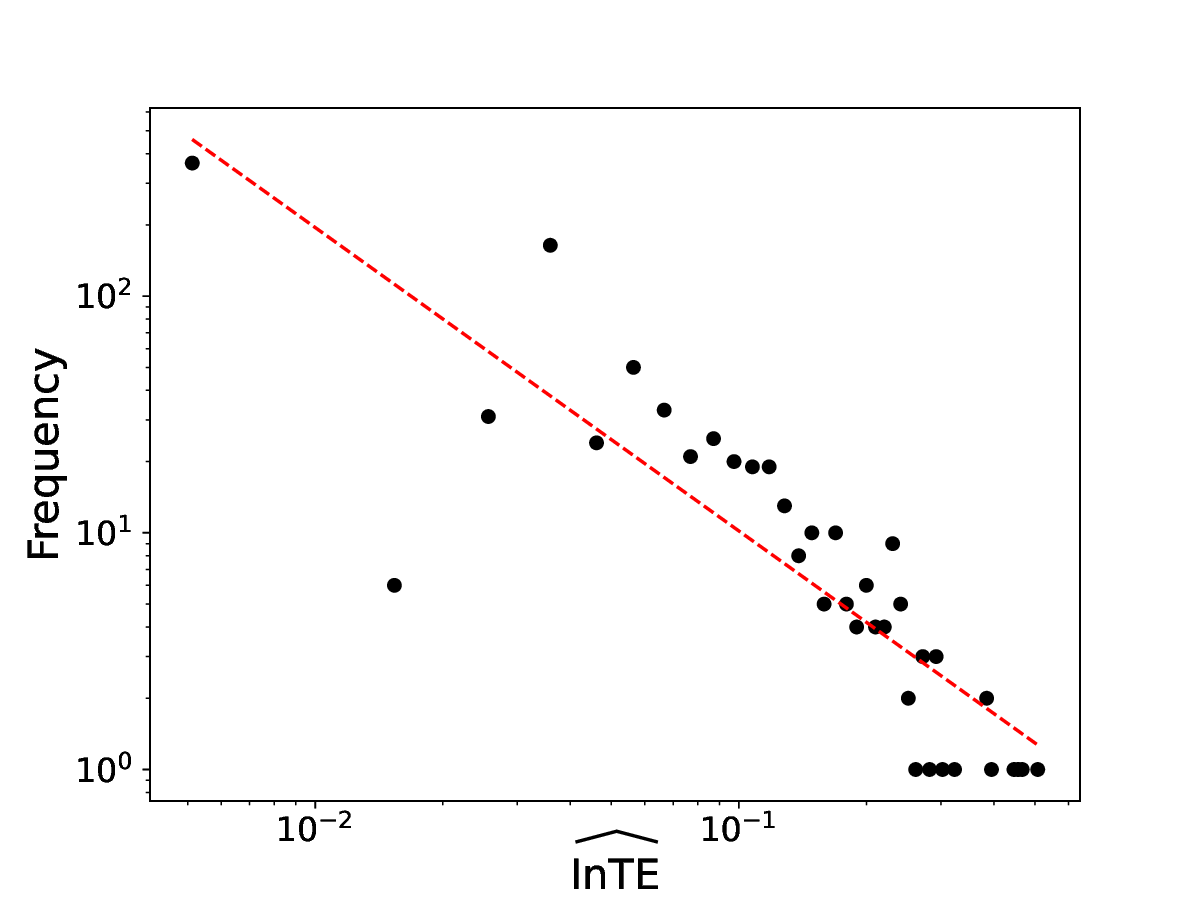}
\caption{Histrogram of $\widehat {\rm INTE}$ (black dots) of OTUs in the saliva microbiota data, along with a power law fit (red dashed line)}
\label{histin} 
\end{figure}

We now apply the current method to microbiota data obtained from a saliva sample observed over 226 days, consisting of 879 nodes, where each node represents an OTU~\cite{2series}. The values of $\widehat {\rm OutTE}$ and $\widehat {\rm InTE}$ for all nodes are compared, with and without pruning, in Figs.~\ref{baout} and ~\ref{bain}, respectively, where the OTUs are sorted in descending order of the pruned estimates, and the top 100 are shown. We observe that $\widehat {\rm OutTE}=0$ for all OTUs without pruning, as mentioned earlier, whereas the values obtained with pruning vary between 0 and 0.9~(Fig.~\ref{baout}). In the case of $\widehat {\rm InTE}$, the estimated values are non-zero even without pruning. However, while the estimates obtained with pruning range between 0 and 0.5, those obtained without pruning can be as high as 3, and are unrelated to those obtained with pruning (Fig.~\ref{bain}), suggesting large estimation errors without pruning. Also, compared to $\widehat {\rm OutTE}$, where a few OTUs have very high values, such peaks are less pronounced for $\widehat {\rm InTE}$, where no values  exceed 0.5. This difference in behavior can also be seen in the histograms, where 50 bins were used to count frequency of occurrence. The histogram of $\widehat {\rm OutTE}$ exhibits a power-law behavior (Fig.~\ref{histout}), whereas that of $\widehat {\rm InTE}$ fits the power law less well~(Fig.~\ref{histin}). 

The top ten OTUs ranked by $\widehat {\rm OutTE}$ and $\widehat {\rm InTE}$ are shown in the table I. Some microbiota in the data are specified only at higher taxonomic levels. We find that two bacteria, {\it Corynebacterium durum} and {\it Fusobacterium}, stand out in terms of high $\widehat {\rm OutTE}$ values. {\it C. durum} is indeed known to play a crucial role in the community of oral microbiota. As a gram-positive bacterium, it is a prolific biofilm and extracellular matrix producer~\cite{durum1}, and a decrease in this bacterium is associated with a disease~\cite{durum2}. {\it Fusobacterium}, particularly {\it Fusobacterium nucleatum}, is also well known as a key player in the community of oral bacteria~\cite{fuso0, fuso3}. As a gram-negative bacterium, {\it F. nucleatum} is a major coaggregation bridge organism linking early and late colonizers in dental biofilm~\cite{fuso1}, and plays a role in carcinogenesis~\cite{fuso2,fuso4}. No particular bacterium stands out in terms of $\widehat {\rm InTE}$ values. Furthermore, InTE measures vulnerability rather than influence, and it is unclear whether vulnerable species can be easily identified experimentally. Therefore, the biological relevance of the top microbiota in terms of $\widehat {\rm InTE}$ is less evident than for $\widehat {\rm OutTE}$.

\begin{table}[h!]
\centering
\begin{tabular}{|c|c|c|c|c|}
\hline
Rank & \multicolumn{2}{c|}{OutTE} & \multicolumn{2}{c|}{InTE} \\
\hline
1 & Corynebacterium Durum & 0.90 & Gemellales (order)\footnote{Families other than Gemellaceae.} & 0.51\\
\hline
2 & Fusobacterium (genus)\footnote{Includes all the OTUs} & 0.90 & Oribacterium (genus)\footnote{Includes all the OTUs}  & 0.47 \\
\hline
3 & Prevotella melaninogenica & 0.62 & Rothia mucilaginosa & 0.45 \\
\hline
4 & Unknown & 0.62 & SR1 (phylum)\footnote{Includes all the classes} & 0.45 \\
\hline
5 & Coriobacteriaceae (family)\footnote{Genera other than Adlercreutzia, Atopobium, Collinsella, Eggerthella, Slackia, and Rubrobacter.} & 0.52 & Flavobacterium succinicans & 0.39 \\
\hline
\end{tabular}
\label{table1}
\caption{Top ten microbiota with highest values of $\widehat {\rm OutTE}$ and $\widehat {\rm InTE}$.}
\end{table}

\section{Conclusion}
In this study, I introduced outgoing and incoming transfer entropy (OutTE and InTE) in interaction networks, as novel measures for quantifying causal influence of each component on the entire system and vice versa, aiming to identify the most influential (hub) and vulnerable (anti-hub) nodes. The new estimation method proposed here, which incorporates a pruning step to exclude unrelated nodes, significantly reduces estimation errors, particularly in scenarios where the number of variables approaches or exceeds the number of samples. 

Through simulations using synthetic data, I demonstrated the effectiveness of the pruning-enhanced method in accurately estimating the OutTE and InTE values. These results suggest that the method can reliably highlight key nodes in a network, providing valuable insights into the system's dynamics. The application of this method to microbiota data from human saliva further validated its utility, successfully identifying bacterial species known to play critical roles in the oral microbiota community. Specifically, the identification of {\it Corynebacterium durum} and {\it Fusobacterium} as key players underscores the biological relevance of the OutTE measure in complex biological systems. The proposed method also has potential applications across various fields, including biology, neuroscience, and social sciences. 

The construction of binary network of meaningful causal relations is a crucial first step for pruning in this approach, where a publicly available state-of-the-art tool has been used~\cite{lizi19, woll19}. Surely, accurate reconstruction of such a network is important for the successful estimation of OutTE and InTE, and possible future development of more fast and accurate causal network reconstruction algorithm will be very helpful for the application of the estimation method presented in this work. Additionally, applying this method to larger and more diverse datasets could further validate its utility and uncover new insights into the structure and function of complex systems.

\begin{acknowledgments} 
This work was supported by the National Research Foundation of Korea, funded by the Ministry of Science and ICT (NRF-2020R1A2C1005956). 
\end{acknowledgments}
\bibliography{TE}

\begin{thebibliography}{41}%
\makeatletter
\providecommand \@ifxundefined [1]{%
 \@ifx{#1\undefined}
}%
\providecommand \@ifnum [1]{%
 \ifnum #1\expandafter \@firstoftwo
 \else \expandafter \@secondoftwo
 \fi
}%
\providecommand \@ifx [1]{%
 \ifx #1\expandafter \@firstoftwo
 \else \expandafter \@secondoftwo
 \fi
}%
\providecommand \natexlab [1]{#1}%
\providecommand \enquote  [1]{``#1''}%
\providecommand \bibnamefont  [1]{#1}%
\providecommand \bibfnamefont [1]{#1}%
\providecommand \citenamefont [1]{#1}%
\providecommand \href@noop [0]{\@secondoftwo}%
\providecommand \href [0]{\begingroup \@sanitize@url \@href}%
\providecommand \@href[1]{\@@startlink{#1}\@@href}%
\providecommand \@@href[1]{\endgroup#1\@@endlink}%
\providecommand \@sanitize@url [0]{\catcode `\\12\catcode `\$12\catcode `\&12\catcode `\#12\catcode `\^12\catcode `\_12\catcode `\%12\relax}%
\providecommand \@@startlink[1]{}%
\providecommand \@@endlink[0]{}%
\providecommand \url  [0]{\begingroup\@sanitize@url \@url }%
\providecommand \@url [1]{\endgroup\@href {#1}{\urlprefix }}%
\providecommand \urlprefix  [0]{URL }%
\providecommand \Eprint [0]{\href }%
\providecommand \doibase [0]{https://doi.org/}%
\providecommand \selectlanguage [0]{\@gobble}%
\providecommand \bibinfo  [0]{\@secondoftwo}%
\providecommand \bibfield  [0]{\@secondoftwo}%
\providecommand \translation [1]{[#1]}%
\providecommand \BibitemOpen [0]{}%
\providecommand \bibitemStop [0]{}%
\providecommand \bibitemNoStop [0]{.\EOS\space}%
\providecommand \EOS [0]{\spacefactor3000\relax}%
\providecommand \BibitemShut  [1]{\csname bibitem#1\endcsname}%
\let\auto@bib@innerbib\@empty
\bibitem [{\citenamefont {Schreiber}(2000)}]{TE00}%
  \BibitemOpen
  \bibfield  {author} {\bibinfo {author} {\bibfnamefont {T.}~\bibnamefont {Schreiber}},\ }\bibfield  {title} {\bibinfo {title} {Measuring information transfer},\ }\href@noop {} {\bibfield  {journal} {\bibinfo  {journal} {Phys. Rev. Lett.}\ }\textbf {\bibinfo {volume} {85}},\ \bibinfo {pages} {461} (\bibinfo {year} {2000})}\BibitemShut {NoStop}%
\bibitem [{\citenamefont {Pearl}(2000)}]{Pe00}%
  \BibitemOpen
  \bibfield  {author} {\bibinfo {author} {\bibfnamefont {J.}~\bibnamefont {Pearl}},\ }\href@noop {} {\emph {\bibinfo {title} {Causality: Models, Reasoning, and Inference}}}\ (\bibinfo  {publisher} {Cambridge University Press},\ \bibinfo {address} {Cambridge},\ \bibinfo {year} {2000})\BibitemShut {NoStop}%
\bibitem [{\citenamefont {Spirtes}\ \emph {et~al.}(2000)\citenamefont {Spirtes}, \citenamefont {Glymour},\ and\ \citenamefont {Scheines}}]{Sp00}%
  \BibitemOpen
  \bibfield  {author} {\bibinfo {author} {\bibfnamefont {P.}~\bibnamefont {Spirtes}}, \bibinfo {author} {\bibfnamefont {C.}~\bibnamefont {Glymour}},\ and\ \bibinfo {author} {\bibfnamefont {R.}~\bibnamefont {Scheines}},\ }\href@noop {} {\emph {\bibinfo {title} {Causation, Prediction, and Search}}}\ (\bibinfo  {publisher} {The MIT Press},\ \bibinfo {address} {Boston},\ \bibinfo {year} {2000})\BibitemShut {NoStop}%
\bibitem [{\citenamefont {Staniek}\ and\ \citenamefont {Lehnertz}(2008)}]{stan08}%
  \BibitemOpen
  \bibfield  {author} {\bibinfo {author} {\bibfnamefont {M.}~\bibnamefont {Staniek}}\ and\ \bibinfo {author} {\bibfnamefont {K.}~\bibnamefont {Lehnertz}},\ }\bibfield  {title} {\bibinfo {title} {Symbolic transfer entropy},\ }\href@noop {} {\bibfield  {journal} {\bibinfo  {journal} {Phys. Rev. Lett.}\ }\textbf {\bibinfo {volume} {100}},\ \bibinfo {pages} {158101} (\bibinfo {year} {2008})}\BibitemShut {NoStop}%
\bibitem [{\citenamefont {Vejmelka}\ and\ \citenamefont {Palus}(2008)}]{vejm08}%
  \BibitemOpen
  \bibfield  {author} {\bibinfo {author} {\bibfnamefont {M.}~\bibnamefont {Vejmelka}}\ and\ \bibinfo {author} {\bibfnamefont {M.}~\bibnamefont {Palus}},\ }\bibfield  {title} {\bibinfo {title} {Inferring the directionality of coupling with conditional mutual information},\ }\href@noop {} {\bibfield  {journal} {\bibinfo  {journal} {Phys. Rev. E}\ }\textbf {\bibinfo {volume} {77}},\ \bibinfo {pages} {026214} (\bibinfo {year} {2008})}\BibitemShut {NoStop}%
\bibitem [{\citenamefont {Ay}\ and\ \citenamefont {Polani}(2008)}]{ay08}%
  \BibitemOpen
  \bibfield  {author} {\bibinfo {author} {\bibfnamefont {N.}~\bibnamefont {Ay}}\ and\ \bibinfo {author} {\bibfnamefont {D.}~\bibnamefont {Polani}},\ }\bibfield  {title} {\bibinfo {title} {Information flows in causal networks},\ }\href@noop {} {\bibfield  {journal} {\bibinfo  {journal} {Advances in Complex Systems}\ }\textbf {\bibinfo {volume} {11}},\ \bibinfo {pages} {17} (\bibinfo {year} {2008})}\BibitemShut {NoStop}%
\bibitem [{\citenamefont {Vicente}\ \emph {et~al.}(2011)\citenamefont {Vicente}, \citenamefont {Wibral}, \citenamefont {Lindner},\ and\ \citenamefont {Pipa}}]{vice11}%
  \BibitemOpen
  \bibfield  {author} {\bibinfo {author} {\bibfnamefont {R.}~\bibnamefont {Vicente}}, \bibinfo {author} {\bibfnamefont {M.}~\bibnamefont {Wibral}}, \bibinfo {author} {\bibfnamefont {M.}~\bibnamefont {Lindner}},\ and\ \bibinfo {author} {\bibfnamefont {G.}~\bibnamefont {Pipa}},\ }\bibfield  {title} {\bibinfo {title} {Transfer entropy—a model-free measure of effective connectivity for the neurosciences},\ }\href@noop {} {\bibfield  {journal} {\bibinfo  {journal} {Journal of Computational Neuroscience}\ }\textbf {\bibinfo {volume} {30}},\ \bibinfo {pages} {45} (\bibinfo {year} {2011})}\BibitemShut {NoStop}%
\bibitem [{\citenamefont {Wibral}\ \emph {et~al.}(2013)\citenamefont {Wibral}, \citenamefont {Pampu}, \citenamefont {Priesemann}, \citenamefont {Siebenhühner}, \citenamefont {Seiwert}, \citenamefont {Lindner}, \citenamefont {Lizier},\ and\ \citenamefont {Vicente}}]{wibr13}%
  \BibitemOpen
  \bibfield  {author} {\bibinfo {author} {\bibfnamefont {M.}~\bibnamefont {Wibral}}, \bibinfo {author} {\bibfnamefont {N.}~\bibnamefont {Pampu}}, \bibinfo {author} {\bibfnamefont {V.}~\bibnamefont {Priesemann}}, \bibinfo {author} {\bibfnamefont {F.}~\bibnamefont {Siebenhühner}}, \bibinfo {author} {\bibfnamefont {H.}~\bibnamefont {Seiwert}}, \bibinfo {author} {\bibfnamefont {M.}~\bibnamefont {Lindner}}, \bibinfo {author} {\bibfnamefont {J.~T.}\ \bibnamefont {Lizier}},\ and\ \bibinfo {author} {\bibfnamefont {R.}~\bibnamefont {Vicente}},\ }\bibfield  {title} {\bibinfo {title} {Measuring information-transfer delays},\ }\href@noop {} {\bibfield  {journal} {\bibinfo  {journal} {PloS ONE}\ }\textbf {\bibinfo {volume} {8}},\ \bibinfo {pages} {e55809} (\bibinfo {year} {2013})}\BibitemShut {NoStop}%
\bibitem [{\citenamefont {Song}\ \emph {et~al.}(2024)\citenamefont {Song}, \citenamefont {Jeong}, \citenamefont {de~los Reyes}, \citenamefont {Lim}, \citenamefont {Cho}, \citenamefont {Yeom}, \citenamefont {Lee}, \citenamefont {Lee}, \citenamefont {Lee},\ and\ \citenamefont {Kim}}]{jae1}%
  \BibitemOpen
  \bibfield  {author} {\bibinfo {author} {\bibfnamefont {Y.~M.}\ \bibnamefont {Song}}, \bibinfo {author} {\bibfnamefont {J.}~\bibnamefont {Jeong}}, \bibinfo {author} {\bibfnamefont {A.~A.~V.}\ \bibnamefont {de~los Reyes}}, \bibinfo {author} {\bibfnamefont {D.}~\bibnamefont {Lim}}, \bibinfo {author} {\bibfnamefont {C.-H.}\ \bibnamefont {Cho}}, \bibinfo {author} {\bibfnamefont {J.~W.}\ \bibnamefont {Yeom}}, \bibinfo {author} {\bibfnamefont {T.}~\bibnamefont {Lee}}, \bibinfo {author} {\bibfnamefont {J.-B.}\ \bibnamefont {Lee}}, \bibinfo {author} {\bibfnamefont {H.-J.}\ \bibnamefont {Lee}},\ and\ \bibinfo {author} {\bibfnamefont {J.~K.}\ \bibnamefont {Kim}},\ }\bibfield  {title} {\bibinfo {title} {Causal dynamics of sleep, circadian rhythm, and mood symptoms in patients with major depression and bipolar disorder: insights from longitudinal wearable device data},\ }\href@noop {} {\bibfield  {journal} {\bibinfo  {journal} {eBioMedicine}\ }\textbf {\bibinfo {volume} {103}},\ \bibinfo {pages} {105094} (\bibinfo
  {year} {2024})}\BibitemShut {NoStop}%
\bibitem [{\citenamefont {Park}\ \emph {et~al.}(2023)\citenamefont {Park}, \citenamefont {Ha},\ and\ \citenamefont {Kim}}]{jae2}%
  \BibitemOpen
  \bibfield  {author} {\bibinfo {author} {\bibfnamefont {S.~H.}\ \bibnamefont {Park}}, \bibinfo {author} {\bibfnamefont {S.}~\bibnamefont {Ha}},\ and\ \bibinfo {author} {\bibfnamefont {J.~K.}\ \bibnamefont {Kim}},\ }\bibfield  {title} {\bibinfo {title} {A general model-based causal inference method overcomes the curse of synchrony and indirect effect},\ }\href@noop {} {\bibfield  {journal} {\bibinfo  {journal} {Nature Communications}\ }\textbf {\bibinfo {volume} {14}},\ \bibinfo {pages} {4287} (\bibinfo {year} {2023})}\BibitemShut {NoStop}%
\bibitem [{\citenamefont {Runge}\ \emph {et~al.}(2012{\natexlab{a}})\citenamefont {Runge}, \citenamefont {Heitzig}, \citenamefont {Marwan},\ and\ \citenamefont {Kurths}}]{runge12a}%
  \BibitemOpen
  \bibfield  {author} {\bibinfo {author} {\bibfnamefont {J.}~\bibnamefont {Runge}}, \bibinfo {author} {\bibfnamefont {J.}~\bibnamefont {Heitzig}}, \bibinfo {author} {\bibfnamefont {N.}~\bibnamefont {Marwan}},\ and\ \bibinfo {author} {\bibfnamefont {J.}~\bibnamefont {Kurths}},\ }\bibfield  {title} {\bibinfo {title} {Quantifying causal coupling strength: A lag-specific measure for multivariate time series related to transfer entropy},\ }\href@noop {} {\bibfield  {journal} {\bibinfo  {journal} {Phys. Rev. E}\ }\textbf {\bibinfo {volume} {86}},\ \bibinfo {pages} {061121} (\bibinfo {year} {2012}{\natexlab{a}})}\BibitemShut {NoStop}%
\bibitem [{\citenamefont {Runge}\ \emph {et~al.}(2012{\natexlab{b}})\citenamefont {Runge}, \citenamefont {Heitzig}, \citenamefont {Petoukhov},\ and\ \citenamefont {Kurths}}]{runge12b}%
  \BibitemOpen
  \bibfield  {author} {\bibinfo {author} {\bibfnamefont {J.}~\bibnamefont {Runge}}, \bibinfo {author} {\bibfnamefont {J.}~\bibnamefont {Heitzig}}, \bibinfo {author} {\bibfnamefont {V.}~\bibnamefont {Petoukhov}},\ and\ \bibinfo {author} {\bibfnamefont {J.}~\bibnamefont {Kurths}},\ }\bibfield  {title} {\bibinfo {title} {Escaping the curse of dimensionality in estimating multivariate transfer entropy},\ }\href@noop {} {\bibfield  {journal} {\bibinfo  {journal} {Phys. Rev. Lett.}\ }\textbf {\bibinfo {volume} {108}},\ \bibinfo {pages} {258701} (\bibinfo {year} {2012}{\natexlab{b}})}\BibitemShut {NoStop}%
\bibitem [{\citenamefont {Sun}\ and\ \citenamefont {Bollt}(2014)}]{sun14}%
  \BibitemOpen
  \bibfield  {author} {\bibinfo {author} {\bibfnamefont {J.}~\bibnamefont {Sun}}\ and\ \bibinfo {author} {\bibfnamefont {E.}~\bibnamefont {Bollt}},\ }\bibfield  {title} {\bibinfo {title} {Causation entropy identifies indirect influences, dominance of neighbors and anticipatory couplings},\ }\href@noop {} {\bibfield  {journal} {\bibinfo  {journal} {Physica D}\ }\textbf {\bibinfo {volume} {267}},\ \bibinfo {pages} {49} (\bibinfo {year} {2014})}\BibitemShut {NoStop}%
\bibitem [{\citenamefont {Sun}\ \emph {et~al.}(2015)\citenamefont {Sun}, \citenamefont {Taylor},\ and\ \citenamefont {Bollt}}]{sun15}%
  \BibitemOpen
  \bibfield  {author} {\bibinfo {author} {\bibfnamefont {J.}~\bibnamefont {Sun}}, \bibinfo {author} {\bibfnamefont {D.}~\bibnamefont {Taylor}},\ and\ \bibinfo {author} {\bibfnamefont {E.}~\bibnamefont {Bollt}},\ }\bibfield  {title} {\bibinfo {title} {Causal network inference by optimal causation entropy},\ }\href@noop {} {\bibfield  {journal} {\bibinfo  {journal} {SIAM J. Appl. Dyn. Syst.}\ }\textbf {\bibinfo {volume} {14}},\ \bibinfo {pages} {73} (\bibinfo {year} {2015})}\BibitemShut {NoStop}%
\bibitem [{\citenamefont {Runge}\ \emph {et~al.}(2015{\natexlab{a}})\citenamefont {Runge}, \citenamefont {Donner},\ and\ \citenamefont {Kurths}}]{runge15a}%
  \BibitemOpen
  \bibfield  {author} {\bibinfo {author} {\bibfnamefont {J.}~\bibnamefont {Runge}}, \bibinfo {author} {\bibfnamefont {R.~V.}\ \bibnamefont {Donner}},\ and\ \bibinfo {author} {\bibfnamefont {J.}~\bibnamefont {Kurths}},\ }\bibfield  {title} {\bibinfo {title} {Optimal model-free prediction from multivariate time series},\ }\href@noop {} {\bibfield  {journal} {\bibinfo  {journal} {Phys. Rev. E}\ }\textbf {\bibinfo {volume} {91}},\ \bibinfo {pages} {052909} (\bibinfo {year} {2015}{\natexlab{a}})}\BibitemShut {NoStop}%
\bibitem [{\citenamefont {Runge}\ \emph {et~al.}(2015{\natexlab{b}})\citenamefont {Runge}, \citenamefont {Petoukhov}, \citenamefont {Donges}, \citenamefont {Hlinka}, \citenamefont {Jajcay}, \citenamefont {Vejmelka}, \citenamefont {Hartman}, \citenamefont {Marwan}, \citenamefont {Palus},\ and\ \citenamefont {Kurths}}]{runge15b}%
  \BibitemOpen
  \bibfield  {author} {\bibinfo {author} {\bibfnamefont {J.}~\bibnamefont {Runge}}, \bibinfo {author} {\bibfnamefont {V.}~\bibnamefont {Petoukhov}}, \bibinfo {author} {\bibfnamefont {J.~F.}\ \bibnamefont {Donges}}, \bibinfo {author} {\bibfnamefont {J.}~\bibnamefont {Hlinka}}, \bibinfo {author} {\bibfnamefont {N.}~\bibnamefont {Jajcay}}, \bibinfo {author} {\bibfnamefont {M.}~\bibnamefont {Vejmelka}}, \bibinfo {author} {\bibfnamefont {D.}~\bibnamefont {Hartman}}, \bibinfo {author} {\bibfnamefont {N.}~\bibnamefont {Marwan}}, \bibinfo {author} {\bibfnamefont {M.}~\bibnamefont {Palus}},\ and\ \bibinfo {author} {\bibfnamefont {J.}~\bibnamefont {Kurths}},\ }\bibfield  {title} {\bibinfo {title} {Identifying causal gateways and mediators in complex spatio-temporal systems},\ }\href@noop {} {\bibfield  {journal} {\bibinfo  {journal} {Nat. Commun.}\ }\textbf {\bibinfo {volume} {6}},\ \bibinfo {pages} {8502} (\bibinfo {year} {2015}{\natexlab{b}})}\BibitemShut {NoStop}%
\bibitem [{\citenamefont {Runge}(2018)}]{runge18}%
  \BibitemOpen
  \bibfield  {author} {\bibinfo {author} {\bibfnamefont {J.}~\bibnamefont {Runge}},\ }\bibfield  {title} {\bibinfo {title} {Causal network reconstruction from time series: From theoretical assumptions to practical estimation},\ }\href@noop {} {\bibfield  {journal} {\bibinfo  {journal} {Chaos}\ }\textbf {\bibinfo {volume} {28}},\ \bibinfo {pages} {075310} (\bibinfo {year} {2018})}\BibitemShut {NoStop}%
\bibitem [{\citenamefont {Runge}\ \emph {et~al.}(2019)\citenamefont {Runge}, \citenamefont {Bathiany}, \citenamefont {Bollt}, \citenamefont {Camps-Valls}, \citenamefont {Coumou}, \citenamefont {Deyle}, \citenamefont {Glymour}, \citenamefont {Kretschmer}, \citenamefont {Mahecha}, \citenamefont {Muñoz-Marí}, \citenamefont {van Nes}, \citenamefont {Peters}, \citenamefont {Quax}, \citenamefont {Reichstein}, \citenamefont {Scheffer}, \citenamefont {Schölkopf}, \citenamefont {Spirtes}, \citenamefont {Sugihara}, \citenamefont {Sun}, \citenamefont {Zhang},\ and\ \citenamefont {Zscheischler}}]{runge19}%
  \BibitemOpen
  \bibfield  {author} {\bibinfo {author} {\bibfnamefont {J.}~\bibnamefont {Runge}}, \bibinfo {author} {\bibfnamefont {S.}~\bibnamefont {Bathiany}}, \bibinfo {author} {\bibfnamefont {E.}~\bibnamefont {Bollt}}, \bibinfo {author} {\bibfnamefont {G.}~\bibnamefont {Camps-Valls}}, \bibinfo {author} {\bibfnamefont {D.}~\bibnamefont {Coumou}}, \bibinfo {author} {\bibfnamefont {E.}~\bibnamefont {Deyle}}, \bibinfo {author} {\bibfnamefont {C.}~\bibnamefont {Glymour}}, \bibinfo {author} {\bibfnamefont {M.}~\bibnamefont {Kretschmer}}, \bibinfo {author} {\bibfnamefont {M.~D.}\ \bibnamefont {Mahecha}}, \bibinfo {author} {\bibfnamefont {J.}~\bibnamefont {Muñoz-Marí}}, \bibinfo {author} {\bibfnamefont {E.~H.}\ \bibnamefont {van Nes}}, \bibinfo {author} {\bibfnamefont {J.}~\bibnamefont {Peters}}, \bibinfo {author} {\bibfnamefont {R.}~\bibnamefont {Quax}}, \bibinfo {author} {\bibfnamefont {M.}~\bibnamefont {Reichstein}}, \bibinfo {author} {\bibfnamefont {M.}~\bibnamefont {Scheffer}}, \bibinfo {author} {\bibfnamefont
  {B.}~\bibnamefont {Schölkopf}}, \bibinfo {author} {\bibfnamefont {P.}~\bibnamefont {Spirtes}}, \bibinfo {author} {\bibfnamefont {G.}~\bibnamefont {Sugihara}}, \bibinfo {author} {\bibfnamefont {J.}~\bibnamefont {Sun}}, \bibinfo {author} {\bibfnamefont {K.}~\bibnamefont {Zhang}},\ and\ \bibinfo {author} {\bibfnamefont {J.}~\bibnamefont {Zscheischler}},\ }\bibfield  {title} {\bibinfo {title} {Inferring causation from time series in earth system sciences},\ }\href@noop {} {\bibfield  {journal} {\bibinfo  {journal} {Nature Communications}\ }\textbf {\bibinfo {volume} {10}},\ \bibinfo {pages} {2553} (\bibinfo {year} {2019})}\BibitemShut {NoStop}%
\bibitem [{\citenamefont {Novelli}\ \emph {et~al.}(2019)\citenamefont {Novelli}, \citenamefont {Wollstadt}, \citenamefont {Mediano}, \citenamefont {Wibral},\ and\ \citenamefont {Lizier}}]{lizi19}%
  \BibitemOpen
  \bibfield  {author} {\bibinfo {author} {\bibfnamefont {L.}~\bibnamefont {Novelli}}, \bibinfo {author} {\bibfnamefont {P.}~\bibnamefont {Wollstadt}}, \bibinfo {author} {\bibfnamefont {P.}~\bibnamefont {Mediano}}, \bibinfo {author} {\bibfnamefont {M.}~\bibnamefont {Wibral}},\ and\ \bibinfo {author} {\bibfnamefont {J.~T.}\ \bibnamefont {Lizier}},\ }\bibfield  {title} {\bibinfo {title} {Large-scale directed network inference with multivariate transfer entropy and hierarchical statistical testing},\ }\href@noop {} {\bibfield  {journal} {\bibinfo  {journal} {Network Neuroscience}\ }\textbf {\bibinfo {volume} {3}},\ \bibinfo {pages} {827} (\bibinfo {year} {2019})}\BibitemShut {NoStop}%
\bibitem [{\citenamefont {Novelli}\ and\ \citenamefont {Lizier}(2021)}]{NL21}%
  \BibitemOpen
  \bibfield  {author} {\bibinfo {author} {\bibfnamefont {L.}~\bibnamefont {Novelli}}\ and\ \bibinfo {author} {\bibfnamefont {J.~T.}\ \bibnamefont {Lizier}},\ }\bibfield  {title} {\bibinfo {title} {{Inferring network properties from time series using transfer entropy and mutual information: Validation of multivariate versus bivariate approaches}},\ }\href@noop {} {\bibfield  {journal} {\bibinfo  {journal} {Network Neuroscience}\ }\textbf {\bibinfo {volume} {5}},\ \bibinfo {pages} {373} (\bibinfo {year} {2021})}\BibitemShut {NoStop}%
\bibitem [{\citenamefont {J.~T.Novelli}\ \emph {et~al.}(2020)\citenamefont {J.~T.Novelli}, \citenamefont {Atay}, \citenamefont {Jost},\ and\ \citenamefont {Lizier}}]{nove20}%
  \BibitemOpen
  \bibfield  {author} {\bibinfo {author} {\bibfnamefont {L.}~\bibnamefont {J.~T.Novelli}}, \bibinfo {author} {\bibfnamefont {F.~M.}\ \bibnamefont {Atay}}, \bibinfo {author} {\bibfnamefont {J.}~\bibnamefont {Jost}},\ and\ \bibinfo {author} {\bibfnamefont {J.~T.}\ \bibnamefont {Lizier}},\ }\bibfield  {title} {\bibinfo {title} {Deriving pairwise transfer entropy from network structure and motifs},\ }\href@noop {} {\bibfield  {journal} {\bibinfo  {journal} {Proc. R. Soc. A}\ }\textbf {\bibinfo {volume} {476}},\ \bibinfo {pages} {20190779} (\bibinfo {year} {2020})}\BibitemShut {NoStop}%
\bibitem [{\citenamefont {James}\ \emph {et~al.}(2016)\citenamefont {James}, \citenamefont {Barnett},\ and\ \citenamefont {Crutchfield}}]{tecrit}%
  \BibitemOpen
  \bibfield  {author} {\bibinfo {author} {\bibfnamefont {R.~G.}\ \bibnamefont {James}}, \bibinfo {author} {\bibfnamefont {N.}~\bibnamefont {Barnett}},\ and\ \bibinfo {author} {\bibfnamefont {J.~P.}\ \bibnamefont {Crutchfield}},\ }\bibfield  {title} {\bibinfo {title} {Information flows? a critique of transfer entropies},\ }\href@noop {} {\bibfield  {journal} {\bibinfo  {journal} {Phys. Rev. Lett.}\ }\textbf {\bibinfo {volume} {116}},\ \bibinfo {pages} {238701} (\bibinfo {year} {2016})}\BibitemShut {NoStop}%
\bibitem [{\citenamefont {Orlandi}\ \emph {et~al.}(2014)\citenamefont {Orlandi}, \citenamefont {Stetter}, \citenamefont {Soriano}, \citenamefont {Geisel},\ and\ \citenamefont {Battaglia}}]{neu1}%
  \BibitemOpen
  \bibfield  {author} {\bibinfo {author} {\bibfnamefont {J.~G.}\ \bibnamefont {Orlandi}}, \bibinfo {author} {\bibfnamefont {O.}~\bibnamefont {Stetter}}, \bibinfo {author} {\bibfnamefont {J.}~\bibnamefont {Soriano}}, \bibinfo {author} {\bibfnamefont {T.}~\bibnamefont {Geisel}},\ and\ \bibinfo {author} {\bibfnamefont {D.}~\bibnamefont {Battaglia}},\ }\bibfield  {title} {\bibinfo {title} {Transfer entropy reconstruction and labeling of neuronal connections from simulated calcium imaging},\ }\href@noop {} {\bibfield  {journal} {\bibinfo  {journal} {PLoS One}\ }\textbf {\bibinfo {volume} {9}},\ \bibinfo {pages} {e98842} (\bibinfo {year} {2014})}\BibitemShut {NoStop}%
\bibitem [{\citenamefont {Wollstadt}\ \emph {et~al.}(2014)\citenamefont {Wollstadt}, \citenamefont {Mart\'{i}nez-Zarzuela}, \citenamefont {Vicente}, \citenamefont {D\'{i}az-Pernas},\ and\ \citenamefont {Wibral}}]{neu2}%
  \BibitemOpen
  \bibfield  {author} {\bibinfo {author} {\bibfnamefont {P.}~\bibnamefont {Wollstadt}}, \bibinfo {author} {\bibfnamefont {M.}~\bibnamefont {Mart\'{i}nez-Zarzuela}}, \bibinfo {author} {\bibfnamefont {R.}~\bibnamefont {Vicente}}, \bibinfo {author} {\bibfnamefont {F.}~\bibnamefont {D\'{i}az-Pernas}},\ and\ \bibinfo {author} {\bibfnamefont {M.}~\bibnamefont {Wibral}},\ }\bibfield  {title} {\bibinfo {title} {Efficient transfer entropy analysis of non-stationary neural time series},\ }\href@noop {} {\bibfield  {journal} {\bibinfo  {journal} {PLoS One}\ }\textbf {\bibinfo {volume} {9}},\ \bibinfo {pages} {e102833} (\bibinfo {year} {2014})}\BibitemShut {NoStop}%
\bibitem [{\citenamefont {Spinney}\ \emph {et~al.}(2017)\citenamefont {Spinney}, \citenamefont {Prokopenko},\ and\ \citenamefont {Lizier}}]{neu3}%
  \BibitemOpen
  \bibfield  {author} {\bibinfo {author} {\bibfnamefont {R.~E.}\ \bibnamefont {Spinney}}, \bibinfo {author} {\bibfnamefont {M.}~\bibnamefont {Prokopenko}},\ and\ \bibinfo {author} {\bibfnamefont {J.~T.}\ \bibnamefont {Lizier}},\ }\bibfield  {title} {\bibinfo {title} {Transfer entropy in continuous time, with applications to jump and neural spiking processes},\ }\href@noop {} {\bibfield  {journal} {\bibinfo  {journal} {Phys. Rev. E}\ }\textbf {\bibinfo {volume} {95}},\ \bibinfo {pages} {032319} (\bibinfo {year} {2017})}\BibitemShut {NoStop}%
\bibitem [{\citenamefont {Kim}\ \emph {et~al.}(2016)\citenamefont {Kim}, \citenamefont {Newth},\ and\ \citenamefont {Christen}}]{soc1}%
  \BibitemOpen
  \bibfield  {author} {\bibinfo {author} {\bibfnamefont {M.}~\bibnamefont {Kim}}, \bibinfo {author} {\bibfnamefont {D.}~\bibnamefont {Newth}},\ and\ \bibinfo {author} {\bibfnamefont {P.}~\bibnamefont {Christen}},\ }\bibfield  {title} {\bibinfo {title} {Macro-level information transfer in social media: Reflections of crowd phenomena},\ }\href@noop {} {\bibfield  {journal} {\bibinfo  {journal} {Neurocomputing}\ }\textbf {\bibinfo {volume} {172}},\ \bibinfo {pages} {84} (\bibinfo {year} {2016})}\BibitemShut {NoStop}%
\bibitem [{\citenamefont {Kim}\ \emph {et~al.}(2021)\citenamefont {Kim}, \citenamefont {Jakobsen}, \citenamefont {Natarajan},\ and\ \citenamefont {Won}}]{tenet}%
  \BibitemOpen
  \bibfield  {author} {\bibinfo {author} {\bibfnamefont {J.}~\bibnamefont {Kim}}, \bibinfo {author} {\bibfnamefont {S.~T.}\ \bibnamefont {Jakobsen}}, \bibinfo {author} {\bibfnamefont {K.~N.}\ \bibnamefont {Natarajan}},\ and\ \bibinfo {author} {\bibfnamefont {K.-J.}\ \bibnamefont {Won}},\ }\bibfield  {title} {\bibinfo {title} {Tenet: gene network reconstruction using transfer entropy reveals key regulatory factors from single cell transcriptomic data},\ }\href@noop {} {\bibfield  {journal} {\bibinfo  {journal} {Nucleic Acids Res.}\ }\textbf {\bibinfo {volume} {49}},\ \bibinfo {pages} {e1} (\bibinfo {year} {2021})}\BibitemShut {NoStop}%
\bibitem [{\citenamefont {Weng}\ \emph {et~al.}()\citenamefont {Weng}, \citenamefont {Kim}, \citenamefont {Natarajan},\ and\ \citenamefont {Won}}]{scmte}%
  \BibitemOpen
  \bibfield  {author} {\bibinfo {author} {\bibfnamefont {G.}~\bibnamefont {Weng}}, \bibinfo {author} {\bibfnamefont {J.}~\bibnamefont {Kim}}, \bibinfo {author} {\bibfnamefont {K.~N.}\ \bibnamefont {Natarajan}},\ and\ \bibinfo {author} {\bibfnamefont {K.-J.}\ \bibnamefont {Won}},\ }\bibfield  {title} {\bibinfo {title} {scm{TE}: multivariate transfer entropy builds interpretable compact gene regulatory networks by reducing false predictions},\ }\href@noop {} {\bibinfo  {journal} {bioRxiv doi: 10.1101/2022.11.08.515579}\ }\BibitemShut {NoStop}%
\bibitem [{\citenamefont {David}\ \emph {et~al.}(2014)\citenamefont {David}, \citenamefont {Materna}, \citenamefont {Friedman}, \citenamefont {Campos-Baptista}, \citenamefont {Blackburn}, \citenamefont {Perrotta}, \citenamefont {Erdman},\ and\ \citenamefont {Alm}}]{2series}%
  \BibitemOpen
\bibfield  {journal} {  }\bibfield  {author} {\bibinfo {author} {\bibfnamefont {L.~A.}\ \bibnamefont {David}}, \bibinfo {author} {\bibfnamefont {A.~C.}\ \bibnamefont {Materna}}, \bibinfo {author} {\bibfnamefont {J.}~\bibnamefont {Friedman}}, \bibinfo {author} {\bibfnamefont {M.~I.}\ \bibnamefont {Campos-Baptista}}, \bibinfo {author} {\bibfnamefont {M.~C.}\ \bibnamefont {Blackburn}}, \bibinfo {author} {\bibfnamefont {A.}~\bibnamefont {Perrotta}}, \bibinfo {author} {\bibfnamefont {S.~E.}\ \bibnamefont {Erdman}},\ and\ \bibinfo {author} {\bibfnamefont {E.~J.}\ \bibnamefont {Alm}},\ }\bibfield  {title} {\bibinfo {title} {Host lifestyle affects human microbiota on daily timescales},\ }\href@noop {} {\bibfield  {journal} {\bibinfo  {journal} {Genome Biology}\ }\textbf {\bibinfo {volume} {15}},\ \bibinfo {pages} {R89} (\bibinfo {year} {2014})}\BibitemShut {NoStop}%
\bibitem [{\citenamefont {Lizier}(2014)}]{jidt}%
  \BibitemOpen
  \bibfield  {author} {\bibinfo {author} {\bibfnamefont {J.~T.}\ \bibnamefont {Lizier}},\ }\bibfield  {title} {\bibinfo {title} {Jidt: An information-theoretic toolkit for studying the dynamics of complex systems},\ }\href@noop {} {\bibfield  {journal} {\bibinfo  {journal} {Front. Robot. AI}\ }\textbf {\bibinfo {volume} {1}},\ \bibinfo {pages} {11} (\bibinfo {year} {2014})}\BibitemShut {NoStop}%
\bibitem [{\citenamefont {Vlachos}\ and\ \citenamefont {Kugiumtzis}(2010)}]{vla10}%
  \BibitemOpen
  \bibfield  {author} {\bibinfo {author} {\bibfnamefont {I.}~\bibnamefont {Vlachos}}\ and\ \bibinfo {author} {\bibfnamefont {D.}~\bibnamefont {Kugiumtzis}},\ }\bibfield  {title} {\bibinfo {title} {Nonuniform state-space reconstruction and coupling detection},\ }\href@noop {} {\bibfield  {journal} {\bibinfo  {journal} {Phys. Rev. E}\ }\textbf {\bibinfo {volume} {82}},\ \bibinfo {pages} {016207} (\bibinfo {year} {2010})}\BibitemShut {NoStop}%
\bibitem [{\citenamefont {Faes}\ \emph {et~al.}(2011)\citenamefont {Faes}, \citenamefont {Nollo},\ and\ \citenamefont {Porta}}]{faes11}%
  \BibitemOpen
  \bibfield  {author} {\bibinfo {author} {\bibfnamefont {L.}~\bibnamefont {Faes}}, \bibinfo {author} {\bibfnamefont {G.}~\bibnamefont {Nollo}},\ and\ \bibinfo {author} {\bibfnamefont {A.}~\bibnamefont {Porta}},\ }\bibfield  {title} {\bibinfo {title} {Transfer entropy—a model-free measure of effective connectivity for the neurosciences},\ }\href@noop {} {\bibfield  {journal} {\bibinfo  {journal} {Journal of Computational Neuroscience}\ }\textbf {\bibinfo {volume} {30}},\ \bibinfo {pages} {45} (\bibinfo {year} {2011})}\BibitemShut {NoStop}%
\bibitem [{\citenamefont {Montalto}\ \emph {et~al.}(2014)\citenamefont {Montalto}, \citenamefont {Faes},\ and\ \citenamefont {Marinazzo}}]{mont14}%
  \BibitemOpen
  \bibfield  {author} {\bibinfo {author} {\bibfnamefont {A.}~\bibnamefont {Montalto}}, \bibinfo {author} {\bibfnamefont {L.}~\bibnamefont {Faes}},\ and\ \bibinfo {author} {\bibfnamefont {D.}~\bibnamefont {Marinazzo}},\ }\bibfield  {title} {\bibinfo {title} {{M}u{TE}: A {MATLAB} toolbox to compare established and novel estimators of the multivariate transfer entropy},\ }\href@noop {} {\bibfield  {journal} {\bibinfo  {journal} {PLoS ONE}\ }\textbf {\bibinfo {volume} {9}},\ \bibinfo {pages} {e109462} (\bibinfo {year} {2014})}\BibitemShut {NoStop}%
\bibitem [{\citenamefont {Wollstadt}\ \emph {et~al.}(2019)\citenamefont {Wollstadt}, \citenamefont {Lizier}, \citenamefont {Vicente}, \citenamefont {Finn}, \citenamefont {Martínez-Zarzuela}, \citenamefont {Mediano}, \citenamefont {Novelli},\ and\ \citenamefont {Wibral}}]{woll19}%
  \BibitemOpen
  \bibfield  {author} {\bibinfo {author} {\bibfnamefont {P.}~\bibnamefont {Wollstadt}}, \bibinfo {author} {\bibfnamefont {J.~T.}\ \bibnamefont {Lizier}}, \bibinfo {author} {\bibfnamefont {R.}~\bibnamefont {Vicente}}, \bibinfo {author} {\bibfnamefont {C.}~\bibnamefont {Finn}}, \bibinfo {author} {\bibfnamefont {M.}~\bibnamefont {Martínez-Zarzuela}}, \bibinfo {author} {\bibfnamefont {P.}~\bibnamefont {Mediano}}, \bibinfo {author} {\bibfnamefont {L.}~\bibnamefont {Novelli}},\ and\ \bibinfo {author} {\bibfnamefont {M.}~\bibnamefont {Wibral}},\ }\bibfield  {title} {\bibinfo {title} {Idtxl: The information dynamics toolkit xl: a python package for the efficient analysis of multivariate information dynamics in networks},\ }\href@noop {} {\bibfield  {journal} {\bibinfo  {journal} {Open Source Software}\ }\textbf {\bibinfo {volume} {4}},\ \bibinfo {pages} {1081} (\bibinfo {year} {2019})}\BibitemShut {NoStop}%
\bibitem [{\citenamefont {Kreth}\ \emph {et~al.}(2024)\citenamefont {Kreth}, \citenamefont {Helliwell}, \citenamefont {Treerat},\ and\ \citenamefont {Merritt}}]{durum1}%
  \BibitemOpen
  \bibfield  {author} {\bibinfo {author} {\bibfnamefont {J.}~\bibnamefont {Kreth}}, \bibinfo {author} {\bibfnamefont {E.}~\bibnamefont {Helliwell}}, \bibinfo {author} {\bibfnamefont {P.}~\bibnamefont {Treerat}},\ and\ \bibinfo {author} {\bibfnamefont {J.}~\bibnamefont {Merritt}},\ }\bibfield  {title} {\bibinfo {title} {Molecular commensalism: how oral corynebacteria and their extracellular membrane vesicles shape microbiome interactions},\ }\href@noop {} {\bibfield  {journal} {\bibinfo  {journal} {Front. Oral. Health}\ }\textbf {\bibinfo {volume} {5}},\ \bibinfo {pages} {1410786} (\bibinfo {year} {2024})}\BibitemShut {NoStop}%
\bibitem [{\citenamefont {Francavilla}\ \emph {et~al.}(2014)\citenamefont {Francavilla}, \citenamefont {Ercolini}, \citenamefont {Piccolo}, \citenamefont {Vannini}, \citenamefont {Siragusa}, \citenamefont {Filippis}, \citenamefont {Pasquale}, \citenamefont {Cagno}, \citenamefont {Toma}, \citenamefont {Gozzi}, \citenamefont {Serrazanetti}, \citenamefont {Angelis},\ and\ \citenamefont {Gobbetti}}]{durum2}%
  \BibitemOpen
  \bibfield  {author} {\bibinfo {author} {\bibfnamefont {R.}~\bibnamefont {Francavilla}}, \bibinfo {author} {\bibfnamefont {D.}~\bibnamefont {Ercolini}}, \bibinfo {author} {\bibfnamefont {M.}~\bibnamefont {Piccolo}}, \bibinfo {author} {\bibfnamefont {L.}~\bibnamefont {Vannini}}, \bibinfo {author} {\bibfnamefont {S.}~\bibnamefont {Siragusa}}, \bibinfo {author} {\bibfnamefont {F.~D.}\ \bibnamefont {Filippis}}, \bibinfo {author} {\bibfnamefont {I.~D.}\ \bibnamefont {Pasquale}}, \bibinfo {author} {\bibfnamefont {R.~D.}\ \bibnamefont {Cagno}}, \bibinfo {author} {\bibfnamefont {M.~D.}\ \bibnamefont {Toma}}, \bibinfo {author} {\bibfnamefont {G.}~\bibnamefont {Gozzi}}, \bibinfo {author} {\bibfnamefont {D.~I.}\ \bibnamefont {Serrazanetti}}, \bibinfo {author} {\bibfnamefont {M.~D.}\ \bibnamefont {Angelis}},\ and\ \bibinfo {author} {\bibfnamefont {M.}~\bibnamefont {Gobbetti}},\ }\bibfield  {title} {\bibinfo {title} {Salivary microbiota and metabolome associated with celiac disease},\ }\href
  {https://doi.org/10.1128/AEM.00362-14} {\bibfield  {journal} {\bibinfo  {journal} {Applied and Environmental Microbiology}\ }\textbf {\bibinfo {volume} {80}},\ \bibinfo {pages} {3416} (\bibinfo {year} {2014})}\BibitemShut {NoStop}%
\bibitem [{\citenamefont {Brennan}\ and\ \citenamefont {Garrett}(2019)}]{fuso0}%
  \BibitemOpen
  \bibfield  {author} {\bibinfo {author} {\bibfnamefont {C.~A.}\ \bibnamefont {Brennan}}\ and\ \bibinfo {author} {\bibfnamefont {W.~S.}\ \bibnamefont {Garrett}},\ }\bibfield  {title} {\bibinfo {title} {Fusobacterium nucleatum — symbiont, opportunist and oncobacterium},\ }\href@noop {} {\bibfield  {journal} {\bibinfo  {journal} {Nature Reviews Microbiology}\ }\textbf {\bibinfo {volume} {17}},\ \bibinfo {pages} {156} (\bibinfo {year} {2019})}\BibitemShut {NoStop}%
\bibitem [{\citenamefont {Groeger}\ \emph {et~al.}(2022)\citenamefont {Groeger}, \citenamefont {Zhou}, \citenamefont {Ruf},\ and\ \citenamefont {Meyle}}]{fuso3}%
  \BibitemOpen
  \bibfield  {author} {\bibinfo {author} {\bibfnamefont {S.}~\bibnamefont {Groeger}}, \bibinfo {author} {\bibfnamefont {Y.}~\bibnamefont {Zhou}}, \bibinfo {author} {\bibfnamefont {S.}~\bibnamefont {Ruf}},\ and\ \bibinfo {author} {\bibfnamefont {J.}~\bibnamefont {Meyle}},\ }\bibfield  {title} {\bibinfo {title} {Pathogenic mechanisms of fusobacterium nucleatum on oral epithelial cells},\ }\href@noop {} {\bibfield  {journal} {\bibinfo  {journal} {Front. Oral. Health}\ }\textbf {\bibinfo {volume} {3}},\ \bibinfo {pages} {2673} (\bibinfo {year} {2022})}\BibitemShut {NoStop}%
\bibitem [{\citenamefont {Kolenbrander}\ \emph {et~al.}(2010)\citenamefont {Kolenbrander}, \citenamefont {Palmer~Jr}, \citenamefont {Periasamy},\ and\ \citenamefont {Jakubovics}}]{fuso1}%
  \BibitemOpen
  \bibfield  {author} {\bibinfo {author} {\bibfnamefont {P.~E.}\ \bibnamefont {Kolenbrander}}, \bibinfo {author} {\bibfnamefont {R.~J.}\ \bibnamefont {Palmer~Jr}}, \bibinfo {author} {\bibfnamefont {S.}~\bibnamefont {Periasamy}},\ and\ \bibinfo {author} {\bibfnamefont {N.~S.}\ \bibnamefont {Jakubovics}},\ }\bibfield  {title} {\bibinfo {title} {Oral multispecies biofilm development and the key role of cell-cell distance},\ }\href@noop {} {\bibfield  {journal} {\bibinfo  {journal} {Nature Reviews Microbiology}\ }\textbf {\bibinfo {volume} {8}},\ \bibinfo {pages} {471} (\bibinfo {year} {2010})}\BibitemShut {NoStop}%
\bibitem [{\citenamefont {Pignatelli}\ \emph {et~al.}(2023)\citenamefont {Pignatelli}, \citenamefont {Nuccio}, \citenamefont {Piattelli},\ and\ \citenamefont {Curia}}]{fuso2}%
  \BibitemOpen
  \bibfield  {author} {\bibinfo {author} {\bibfnamefont {P.}~\bibnamefont {Pignatelli}}, \bibinfo {author} {\bibfnamefont {F.}~\bibnamefont {Nuccio}}, \bibinfo {author} {\bibfnamefont {A.}~\bibnamefont {Piattelli}},\ and\ \bibinfo {author} {\bibfnamefont {M.}~\bibnamefont {Curia}},\ }\bibfield  {title} {\bibinfo {title} {The role of fusobacterium nucleatum in oral and colorectal carcinogenesis},\ }\href@noop {} {\bibfield  {journal} {\bibinfo  {journal} {Microorganisms}\ }\textbf {\bibinfo {volume} {11}},\ \bibinfo {pages} {2358} (\bibinfo {year} {2023})}\BibitemShut {NoStop}%
\bibitem [{\citenamefont {Gholizadeh}\ \emph {et~al.}(2017)\citenamefont {Gholizadeh}, \citenamefont {Eslami},\ and\ \citenamefont {Kafil}}]{fuso4}%
  \BibitemOpen
  \bibfield  {author} {\bibinfo {author} {\bibfnamefont {P.}~\bibnamefont {Gholizadeh}}, \bibinfo {author} {\bibfnamefont {H.}~\bibnamefont {Eslami}},\ and\ \bibinfo {author} {\bibfnamefont {H.~S.}\ \bibnamefont {Kafil}},\ }\bibfield  {title} {\bibinfo {title} {Carcinogenesis mechanisms of fusobacterium nucleatum},\ }\href {https://doi.org/https://doi.org/10.1016/j.biopha.2017.02.102} {\bibfield  {journal} {\bibinfo  {journal} {Biomedicine and Pharmacotherapy}\ }\textbf {\bibinfo {volume} {89}},\ \bibinfo {pages} {918} (\bibinfo {year} {2017})}\BibitemShut {NoStop}%
\end{thebibliography}%

\appendix
\section{Proof of Nonnegativity of Various Information Theoretic Measures}
Shannon entropy H quantifies uncertainty about the state of a system, defined as:
\begin{equation}
  {\rm H}(X) \equiv  \left\langle -\log P(X) \right\rangle,
\end{equation}
where the random variable $X$ represents the states of the system, and $P(X)$ denotes its probability distribution\footnote{Here the base of the log will be left arbitrary since the results do not depend on it.}. Since $P(X)$ is always less than or equal to one, we have $-\log P(X) \ge 0$, from which we get ${\rm H}(X) = \left\langle - \log P(X) \right\rangle \ge 0$. 

The same argument applies to the conditional entropy $H(Y/X)$, as the conditional probability also satisfies the property of usual probability, $P(Y/X) \le 1 $, which results in:
\begin{equation}
  H(Y/X) \equiv  \left\langle -\log P(Y/X) \right\rangle  \ge 0.
\end{equation}
To prove the non-negativity of mutual information, we use Jensen's inequality, which states that for any convex function $f(x)$ and any weights $p_i\ (i=1, \cdots m)$ (where $\sum_{i=1}^m p_i=1$), we have:
$$\sum_{i=1}^m p_i f(x_i)  \geq f(\sum_{i=1}^m p_i x_i).$$ Applying this inequality to the convex function $f(x) = -\log(x)$, we get:
\begin{eqnarray}
    I(X,Y) &=&  \sum_{x,y} P(X=x,Y=y) \left( -\log \frac{P(X=x)P(Y=y)}{P(X=x,Y=y)} \right)\nonumber\\
    &\geq& -\log \left(  \sum_{x,y} P(X=x,Y=y) \frac{P(X=x)P(Y=y)}{P (X=x,Y=y)} \right) \nonumber\\
    &=& -\log \left(  \sum_{x} P(X=x) \sum_{y}P(Y=y) \right) = 0.\label{minn}
\end{eqnarray}
The same argument applies to conditional mutual information:
\begin{eqnarray}
    I(X,Y/Z) &=&  \sum_z P(Z=z) \sum_{x,y} P(X=x,Y=y/Z=z) \left( -\log \frac{P(X=x/Z=z) P(Y=y/Z=z)}{P(X=x,Y=y/Z=z)} \right)\nonumber\\
    &\geq& 0,
\end{eqnarray}
where the last inequality follows by applying Eq.~(\ref{minn}) for each $z$.

Since transfer entropy and conditional transfer entropy are conditional mutual information, they are all nonnegative. 
\section{Underestimation of Conditional Entropy for Small Sample Sizes}

Conditional entropy \( H(Y/X) \) measures the amount of uncertainty remaining in a random variable \( Y \) given that the value of another random variable \( X \) is known. The true conditional entropy is defined as:
\[
H(Y|X) = -\sum_x P(X=x) \sum_y P(Y=y|X=x) \log P(Y=y|X=x).
\]
When estimating conditional entropy from a finite sample, the empirical conditional entropy \( \hat{H}(Y/X) \) is calculated by substituting the true probabilities with their empirical estimates based on the observed data:
\[
\hat{H}(Y/X) = -\sum_{x } \hat{P}(X=x) \sum_y \hat{P}(X=y|X=x) \log \hat{P}(X=y|X=x),
\]
where empirical probabilities are calculated as:
\[
\hat{P}(X=x) \equiv \frac{N(x)}{N},\quad   \hat{P}(Y=y|X=x) \equiv \frac{N(x, y)}{N(x)}
\]
with \( N(x, y) \) being the number of times the pair \( (x, y) \) occurs, and \( N(x) \) the number of times \( x \) occurs, and $N$ the total number of observations. In practice, the estimated conditional entropy \( \hat{H}(Y|X) \) tends to be smaller than the true entropy \( H(Y|X) \) when the sample size is small. This underestimation occurs due to insufficient observations, which leads to a biased estimation of probabilities.

The primary reason for the underestimation of conditional entropy lies in the concavity of the logarithm function and the bias introduced by finite sample sizes. Note that Jensen's inequality can be written in the form
\begin{equation}
\langle f(X) \rangle \leq f(\langle X \rangle )     
\end{equation}
for any random variable $X$ and a concave function $f(x)$. The logarithm function \( \log(x) \) is concave, which means that for any random variable \( Z \):
\[
\langle \log Z \rangle \leq \log \langle Z \rangle
\]
In the context of conditional entropy, this inequality implies that the expected value of the logarithm of the estimated probability is less than the logarithm of the true probability. When probabilities are estimated from a small sample, the estimates \( \hat{P}(Y=y|X=x) \) are typically more concentrated around zero for rare events, leading to a lower average entropy. That is, for small sample sizes, many possible pairs \( (x, y) \) may not be observed at all, leading to \( \hat{P}(Y=y|X=x) = 0 \) for these pairs. Since \(  \hat{P}(Y=y|X=x) \log \hat{P}(Y=y|X=x) \) is defined as $\lim_{p \to 0} p \log p = 0$ when \( \hat{P}(Y=y|X=x) = 0 \),  this results in a lower estimated entropy. In contrast, the true probability \( P(Y=y|X=x) \) might be nonzero, leading to a nonzero contribution to the true entropy \( H(Y|X) \).

For example, consider a simple case where \( X \) and \( Y \) are binary variables, and the true conditional probabilities are:
\[
P(Y=0 | X=0) = 1/2, \quad P(Y=1 | X=0) = 1/2
\]

\[
P(Y=0 | X=1) = 1/2, \quad P(Y=1 | X=1) = 1/2,
\]
Suppose we have just two observations, with $(X, Y)=(0, 0)$ and $(X, Y)=(1, 1)$.  The empirical conditional probabilities are:
\[
\hat{P}(Y=0 | X=0) = 1, \quad \hat{P}(Y=1 | X=0) = 0
\]
\[
\hat{P}(Y=0 | X=1) = 0, \quad \hat{P}(Y=1 | X=1) = 1
\]
The empirical conditional entropy \( \hat{H}(Y|X)=0 \) calculated from these estimates is less than the true conditional entropy \( H(Y|X)=1 \) calculated from the true probabilities, illustrating the underestimation bias.

\newpage

    
    

\end{document}